\documentclass[prb,twocolumn,showpacs]{revtex4} 
\usepackage{graphicx} 
\usepackage{amsmath} 
\usepackage{amssymb} 
\usepackage{dcolumn} 
\usepackage{float} 
\usepackage{bm} 
\usepackage[breaklinks=true,colorlinks,citecolor=blue,linkcolor=blue,urlcolor=blue]{hyperref}

\DeclareMathAlphabet{\bi}{OML}{cmm}{b}{it}

\def\be{\begin{equation}}
\def\ee{\end{equation}}
\def\bearr{\begin{eqnarray}}
\def\eearr{\end{eqnarray}}
\def\la{\langle}
\def\ra{\rangle}

\def\bs{\boldsymbol}

\begin{document}
\title{Dynamics of a quasiparticle in the $\alpha$-T$_3$ model: Role of
pseudospin polarization and transverse magnetic field on \textbf{\textit{zitterbewegung}}}
\bigskip

\author{Tutul Biswas}
\email{tbtutulm53@gmail.com}
\author{Tarun Kanti Ghosh
}
\email{tkghosh@iitk.ac.in}
\normalsize
\affiliation
{$\color{blue}{^\ast}$Department of Physics, University of North Bengal, Raja Rammohunpur-734013, India\\
$\color{blue}{^\ast}$Department of Physics, Vivekananda Mahavidyalaya, Burdwan-713103, India\\
$\color{blue}{^\dagger}$Department of Physics, Indian Institute of Technology-Kanpur,
Kanpur-208 016, India}
\date{\today}
\begin{abstract} 
We consider the $\alpha$-$T_3$ model which provides a smooth crossover between the honeycomb lattice with pseudospin $1/2$ 
and the dice lattice with pseudospin $1$ through the variation of a parameter $\alpha$.
We study the dynamics of a wave packet representing a quasiparticle in the $\alpha$-T$_3$
model with zero and finite transverse magnetic field. For zero field, it is shown that the wave packet undergoes a
transient $zitterbewegung$ (ZB). Various features of ZB
depending on the initial pseudospin polarization of the wave packet have been revealed.
For an intermediate value of the parameter $\alpha$ i.e. for $0<\alpha<1$ the resulting ZB consists of two distinct
frequencies when the wave packet was located initially
in $rim$ site. However, the wave packet exhibits single frequency ZB for $\alpha=0$ and $\alpha=1$. It is 
also unveiled that the frequency of ZB corresponding to $\alpha=1$ gets exactly half of that corresponding to 
the $\alpha=0$ case. On the other hand, when the initial wave packet was in $hub$ site, the ZB consists 
of only one frequency for all values of $\alpha$. Using stationary phase approximation we find analytical expression
of velocity average which can be used to extract the associated timescale over which the transient nature of
ZB persists. On the contrary the wave packet undergoes permanent ZB in presence of a transverse magnetic field.
Due to the presence of large number of Landau energy levels the oscillations in ZB appear to be much more complicated. The 
oscillation pattern depends significantly on the initial pseudospin polarization of the wave packet. Furthermore,
it is revealed that the number of the frequency components involved in ZB depends on the parameter $\alpha$.
\end{abstract}


\maketitle

\section{Introduction}
The conception $zitterbewegung$ (ZB) stands for an outlandish quantum motion, 
of a Dirac particle in vacuum, having length scale of the order of Compton wave length.
It was originally envisioned by Schr\"{o}dinger in 1930\cite{Schro}. The main obstruction to 
establish the existence of ZB in vacuum experimentally 
is its ultra-short length scale.
However, a ray of hope was shown in 2005 when Zawadki\cite{zawadki} argued that a narrow 
gap semiconductor not only can host the intriguing phenomenon ZB but also the associated 
length scale can be enhanced up to five orders higher than that in vacuum.
As a result, subsequent years witnessed immense interest in ZB in numerous systems\cite{zbgen}
including spin-orbit coupled two 
dimensional (2D) electron/hole gases\cite{john,zb2d1,zb2d2,zb2d3,zb2d4,zb2d5,zb2dH,zb2dH2},
superconductors\cite{zbsup}, sonic crystal\cite{zbsonic}, photonic crystal\cite{zbphoton,zbopp},
carbon nanotube\cite{zbcnt}, graphene\cite{zbgrph1,zbgrph2,zbgrph3,zbgrph4,zbgrph5}, other Dirac
materials\cite{zbtopo,zbMos2,zbMos22} and ultra-cold atomic
gases\cite{zbbec1,zbbec2,zbbec3,zbbec4}.

There exists several understandings behind the cause of ZB. It is  believed that  
ZB happens as an outcome of the interference between positive and negative energy solutions of the Dirac equation.
Huang\cite{huang} put forward a theory to establish a connection between ZB and electron's intrinsic magnetic dipole moment.
Later, Schliemann\cite{zb2d2} $et$ $al$ interpreted that ZB in a quantum well occurs as a consequence of spin rotation due to 
spin-orbit interaction (SOI). It is also mentioned\cite{zbbec1} that the ZB can be interpreted as a measurable aftermath 
of Berry phase in momentum space. Moreover, for a multiband quantum system, an explicit relation between Berry curvature
and amplitude of ZB was also established\cite{zbCsr}.

In general ZB has permanent character i.e. oscillations do not die out in time.
When an electron is illustrated by a wave packet the resulting ZB undergoes a
transient nature according to Lock\cite{lock}. It is also proposed recently\cite{zb2d5} that the permanent
behavior of ZB in a spin-orbit coupled 2D electron gas can be restored by considering a time 
dependent Rashba SOI. 

Furthermore, an intriguing quantum transport related phenomenon such as
minimal conductivity\cite{ZbKats} of graphene was understood in the light 
of ZB. Very recently, Iwasaki {\it et} {\it al}\,\cite{ZbIwa} demonstrated experimentally that 
conductance fluctuations in InAs quantum wells occur as a possible consequence of ZB.

From the perspective of ZB, most of the studies in electronic systems are mainly concerned about 
the ZB of quasiparticles with spin/pseudospin $S=1/2$.
However, there exists an example which portrays ZB of spin-1 ultra-cold atom\cite{zbbec2}.
To the best of our knowledge no such example of ZB of a quasiparticle with spin/pseudospin beyond
$S=1/2$ exists in typical condensed matter
systems. Hence, in this article we consider the ZB effect in a relatively
new model named as $\alpha$-T$_3$ model in which quasiparticles are 
characterized by an enlarged pseudospin $S\geq1/2$. The sole motivation behind adopting this model
is due to growing interest in systems which are described by the generalized Dirac-Weyl 
equation with arbitrary pseudospin $S$ \cite{dice_S1,dice_S2,dice_S3}.
Via the variation of a parameter $\alpha$, the
$\alpha$-T$_3$ model reveals a smooth changeover from graphene ($\alpha=0$) to 
dice or T$_3$ lattice ($\alpha=1$). There has been a lot of studies in T$_3$ 
lattice from the standpoint of topological localization \cite{dice1,dice2}, 
magnetic frustration \cite{frust1,frust2}, Klein tunneling \cite{dice_Klein}, minimal conductivity\cite{minT3},
plasmon\cite{plasm} etc.
Moreover, the existence of the $T_3$-model within the framework of optical lattice\cite{dice_opt} and 
semiconductor structure\cite{dice_grow} is also predicted recently.
In addition, the interest in $\alpha$-$T_3$ model\cite{dice_alph} is growing rapidly nowadays.
According to Malcolm and Nicol \cite{dice_alph2} a Hg$_{1-x}$Cd$_x$Te quantum well can be considered
as $\alpha$-T$_3$ model with $\alpha=1/\sqrt{3}$ at a particular doping. 
The connection of the parameter $\alpha$ with the Berry phase makes the $\alpha$-T$_3$ model
more interesting. In recent years, a number of studies have been performed on this model
within the context of orbital magnetic response\cite{dice_alph}, magneto-transport\cite{dice_MagTr},
optical conductivity\cite{dice_Berry,dice_MagOP1,dice_MagOP2}, quantum tunneling\cite{dice_QT},
Wiess oscillations\cite{dice_Wiess} etc.

In this work we have investigated the problem of ZB of a quasiparticle in $\alpha$-T$_3$ model. 
We choose an initial Gaussian wave packet with definite pseudospin polarization to represent a
quasiparticle. For $0<\alpha<1$, the quasiparticle undergoes ZB which is transient in nature and 
consists of two frequencies, namely, $2\Omega_q$ and $\Omega_q$. The interference between 
conduction and valence band leads to occur ZB with frequency $2\Omega_q$ whereas the 
$\Omega_q$-frequency ZB occurs as a result of interference between either conduction and flat band 
or flat and valence band. The nature of ZB depends significantly on the type of the initial 
pseudospin polarization. Particularly, when the initial pseudospin polarization was along $z$-direction,
various interesting features of ZB emerge. For example, when the initial wave packet was concentrated
entirely in any of the $rim$ sites the resulting ZB has two above mentioned frequencies for a finite $\alpha$.
In this case we reveal a transition from $2\Omega_q$-frequency ZB to $\Omega_q$-frequency ZB as $\alpha$ is 
tuned from $0$ to $1$. When the wave packet was located initially in the $hub$ site, the resulting ZB has 
only one frequency $2\Omega_q$ for all possible values of $\alpha$. In the limit of large width of wave packet
we obtain analytical expression for the expectation value of velocity operator from which one can 
extract the timescale over which the ZB dies out. We also consider the effect of other possible 
pseudospin polarization on ZB. In addition we incorporate the effect of an external quantizing magnetic
field on the ZB. In this case the temporal behavior of ZB is permanent. The effect of different pseudospin
polarization has also been discussed. Using Fourier transformation we obtain frequency components involved in ZB 
for a particular choice of pseudospin polarization. We also find that the number of frequencies
present in ZB depends significantly on the parameter $\alpha$.

Rest of the present paper is organized in the following fashion. In section II, we discuss zero field 
ZB by incorporating the basic informations about physical system, time evolution of wave packet,
and rigorous calculations of the expectation values of physical observables. The effect of perpendicular
magnetic field on ZB is considered in section III. We summarize the obtained results in section IV.

\section{In absence of magnetic field}

\subsection{Lattice structure and low lying energy states}
As depicted in Fig. 1(a), $\alpha$-T$_3$ model has honeycomb lattice structure with an 
additional site at the center of each hexagon. Each unit cell (shown by the dashed rhombus) contains 
$3$ sites.  With respect to the co-ordination number (CN), those sites are 
classified into two categories, namely $rim$ and $hub$
sites. As evident from Fig. 1(a) sites B and C are both $rim$ sites having CN $3$ whereas
site A is known as $hub$ site with CN $6$. Note that each nearest-neighbor pair consists 
of one $rim$ and one $hub$ sites. The sites A and B are connected through hopping parameter
$t$ while hopping energy between A and C is $\alpha t$.

\begin{figure}[h!]
\begin{center}\leavevmode
\includegraphics[width=70mm, height=100mm]{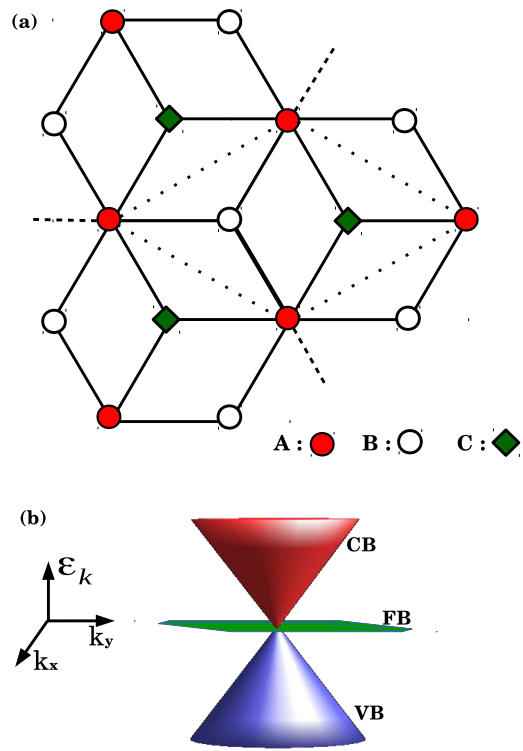}
\caption{(a) The lattice structure of a $\alpha$-T$_3$ model. Here, A(denoted by filled circle) is the 
hub site while B(denoted by empty circle) and C(denoted by solid square) are rim sites. The hopping 
energy between A and B is $t$ whereas the hopping energy between A and C is $\alpha t$. The dotted 
rhombus denotes the unit cell. The dashed lines emerging from A sites indicate that each A site is connected
to the next C site.
(b) The energy surfaces are drawn as a function of wave vector ${\bf k}$ in arbitrary units. The zero energy FB
lies in the ($k_x, k_y$) plane whereas other bands exhibit linear dispersion with ${\bf k}$.}
\end{center}
\end{figure}

The low energy excitations near 
the Dirac point in a particular valley  are described by the following
Hamiltonian\cite{dice_alph}
\begin{eqnarray}\label{HamI}
H({\bf p})=
\begin{pmatrix}
    0      &  f_{\bf p}\cos\varphi &   0 \\
    f_{\bf p}^\ast\cos\varphi  &  0  &  f_{\bf p}\sin\varphi \\
    0     &  f_{\bf p}^\ast\sin\varphi  &  0
\end{pmatrix},
\end{eqnarray}
where $f_{\bf p}=v_F(\zeta p_x-ip_y)$ with $v_F$ being the Fermi velocity. 
The valley index $\zeta$ takes a value $+1(-1)$ for ${\rm K}({\rm K}^\prime)$ valley.
Finally, following relation $\alpha=\tan\varphi$ holds between $\alpha$ and $\varphi$.
The energy spectrum corresponding to the Hamiltonian given in Eq. (\ref{HamI}) consists
of three branches. Out of them two are linearly dispersing
$\varepsilon_{\bf k}^\lambda=\lambda \hbar v_F k$ with $\lambda=\pm1$ 
and $k=\vert {\bf p}\vert/\hbar$, known as conic band. The conic band itself consists of 
conduction band (CB) and valence band (VB) corresponding to $\lambda=+1$ and $\lambda=-1$, respectively.
The other energy branch is dispersionless and is known as flat band (FB).
All these energy branches are depicted in Fig.1(b). The wave functions corresponding
to conic band and FB around the ${\rm K}$-valley are, respectively, given by
\begin{eqnarray}
\psi_{\bf k}^{\lambda}({\bf r})=\frac{1}{\sqrt{2}}\left(
\begin{array}{c}
\cos\varphi\,e^{-i\theta}\\
\lambda\\
\sin\varphi\,e^{i\theta}
\end{array}\right)\frac{e^{i{{\bf k}\cdot {\bf r}}}}{2\pi}
\end{eqnarray}
and 
\begin{eqnarray}
\psi_{\bf k}^{0}({\bf r})=\frac{1}{\sqrt{2}}\left(
\begin{array}{c}
\sin\varphi\,e^{-i\theta}\\
0\\
-\cos\varphi\,e^{i\theta}
\end{array}\right)\frac{e^{i{{\bf k}\cdot {\bf r}}}}{2\pi},
\end{eqnarray}
where $\theta$ is the polar angle of the wave vector ${\bf k}$.

\subsection{Velocity and pseudospin operators}
From Eq. (\ref{HamI}), using the relation $v_i=\partial{H}/\partial{p_i}$ with $i=x,y$, one can obtain
the following components of the velocity matrix for the ${\rm K}$-valley as 
\begin{eqnarray}
v_x=v_F
\begin{pmatrix}
    0      &  \cos\phi &   0 \\
    \cos\phi  &  0  &  \sin\phi \\
    0     &  \sin\phi  &  0
\end{pmatrix}
\end{eqnarray}
and

\begin{eqnarray}
v_y=v_F
\begin{pmatrix}
    0      &  -i\cos\phi &   0 \\
    i\cos\phi  &  0  &  -i\sin\phi \\
    0     &  i\sin\phi  &  0
\end{pmatrix}.
\end{eqnarray}

The $x$ and $y$ components of the pseudospin operator are governed from 
the velocity operators as $S_i=\hbar v_i/v_F$ with $i=x,y$. The $z$-component of ${\bf S}$ is
obtained through the commutation relation $[S_x,S_y]=i\hbar S_z$ as
\begin{eqnarray}
 S_z=2\hbar
\begin{pmatrix}
    \cos^2\phi      &  0 &   0 \\
    0  &  -\cos(2\phi)  &  0 \\
    0     &  0  &  -\sin^2\phi
\end{pmatrix}.
\end{eqnarray}

\subsection{Time evolution of a initial wave packet}
To study the dynamics of a quasiparticle, it is important to 
know the corresponding wave function at a later time $t$. In the following, we consider the initial
wave function representing the quasiparticle to be a plane wave modulated by a Gaussian wave packet
\begin{eqnarray}
 \Psi({\bf r},0)=\frac{1}{2\pi C}\int d{\bf k}\, a({\bf k},0)e^{i{\bf k}\cdot {\bf r}}
 \begin{pmatrix}
    c_1\\
    c_2\\
    c_3
\end{pmatrix},
\end{eqnarray}
where the wave packet $a({\bf k},0)=(d/\sqrt{\pi}) e^{-d^2({\bf k}-{\bf k_0})^2/2}$ is centered 
at some wave vector ${\bf k}={\bf k_0}$ and $d$ is its width. Note that the
wave packet was polarized initially along any arbitrary 
direction characterized by the constants $c_1$, $c_2$, and $c_3$ which are, in general, 
complex numbers. The normalization constant $C$ is defined as
$C=\sqrt{\vert c_1\vert^2+\vert c_2\vert^2+\vert c_3\vert^2}$.
The wave function, at a later time $t$, can be obtained by applying appropriate time evolution 
operator onto $\Psi({\bf r},0)$. We find the time evolved state as
\begin{eqnarray}\label{WaveT}
 \Psi({\bf r},t)=\frac{1}{2\pi}\int d{\bf k}\, a({\bf k},0)e^{i{\bf k}\cdot {\bf r}}
 \begin{pmatrix}
\kappa_1({\bf k},t)\\
\kappa_2({\bf k},t)\\
\kappa_3({\bf k},t)
\end{pmatrix},
\end{eqnarray}
where $\kappa_{\mu}({\bf k},t)$'s, with $\mu=1,2,3$, are obtained from the following matrix equation
\begin{widetext}
\begin{eqnarray}\label{Ham}
 \begin{pmatrix}
\kappa_1({\bf k},t)\\
\kappa_2({\bf k},t)\\
\kappa_3({\bf k},t)
\end{pmatrix}
=\frac{1}{C}
\begin{pmatrix}
\sin^2\varphi+\cos^2\varphi \cos(\Omega_{\bf k} t)  & -i\cos\varphi\,e^{-i\theta}\sin(\Omega_{\bf k}t) &  
\sin\varphi\, \cos\varphi\,e^{-2i\theta}[\cos(\Omega_{\bf k}t)-1]  \\
-i\cos\varphi\, e^{i\theta}\sin(\Omega_{\bf k} t))  &  \cos(\Omega_{\bf k}t)  & 
-i\sin\varphi\,e^{-i\theta}\sin(\Omega_{\bf k}t) \\
 \sin\varphi\, \cos\varphi\, e^{2i\theta}[\cos(\Omega_{\bf k}t)-1] & -i\sin\varphi\, e^{i\theta}\,\sin(\Omega_{\bf k}t) &
 \cos^2\varphi+\sin^2\varphi\, \cos(\Omega_{\bf k} t)
\end{pmatrix}
\begin{pmatrix}
 c_1\\
 c_2\\
 c_3
\end{pmatrix}
\end{eqnarray}
\end{widetext}

It is noteworthy that Eq. (\ref{WaveT}) is the Fourier transformation of the following 
function
\begin{eqnarray}\label{Wave_TM}
 \Phi({\bf k},t)= a({\bf k},0)
 \begin{pmatrix}
    \kappa_1({\bf k},t)\\
   \kappa_2({\bf k},t)\\
   \kappa_3({\bf k},t)
\end{pmatrix}.
\end{eqnarray}
Equation (\ref{Wave_TM}) represents the time evolved state in the momentum space which can be 
used to calculate the expectation values of various physical observables.

\subsection{Expectation value of the velocity operator}
The expectation value of a physical observable corresponding to an operator $\hat{O}$ is defined as 
$\la \hat{O}(t)\ra=\int d{\bf k} \Phi^\dagger({\bf k},t)\hat{O}\Phi({\bf k},t)$. Instead of 
evaluating the expectation value of position operator we prefer here to calculate that of 
velocity operator because the former is always obtained by integrating the later with
suitable initial conditions.

Now the expectation values of the components of velocity operator can be obtained as
\begin{eqnarray}\label{vX}
 \la v_x\ra=2v_F\int d{\bf k} \vert a({\bf k},0)\vert^2 {\rm Re} \big(\cos\varphi\, \kappa_1^\ast \kappa_2 
 +\sin\varphi \, \kappa_2^\ast \kappa_3\big)
\end{eqnarray}
and
\begin{eqnarray}\label{vY}
 \la v_y\ra=2v_F\int d{\bf k} \vert a({\bf k},0)\vert^2 {\rm Im} \big(\cos\varphi\, \kappa_1^\ast \kappa_2 
 +\sin\varphi \, \kappa_2^\ast \kappa_3\big).
\end{eqnarray}

By defining $\Sigma_{\mu \sigma}(t)=\int d{\bf k}\vert a({\bf k},0)\vert^2 \kappa_{\mu}^\ast({\bf k},t) 
\kappa_\sigma({\bf k},t)$
with $\mu,\sigma=1,2,3$, Eqs. (\ref{vX}) and (\ref{vY}) can be further written in the following compact form 
\begin{eqnarray}\label{VeloF}
\begin{pmatrix}
  \la v_x(t)\ra\\
  \la v_y(t)\ra
\end{pmatrix}
=2v_F
\begin{pmatrix}
 {\rm Re}\\
 {\rm Im}
\end{pmatrix}
\Big(\cos\varphi \Sigma_{12}(t)+\sin\varphi \Sigma_{23}(t)\Big).
\end{eqnarray}

Without any loss of generality, we consider the initial wave packet was
moving along +$x$ direction with wave vector ${\bf k_0}=k_0\hat{x}$.
After doing the angular integration, we obtain $\Sigma_{12}(t)$ and $\Sigma_{23}(t)$ as
\begin{widetext}
\begin{eqnarray}\label{Sig1}
&&\Sigma_{12}(t)
=\frac{2}{C^2} e^{-a^2}\int_0^\infty dq\,q\,e^{-q^2}\Bigg[\Big\{ c_1^\ast c_2\sin^2\varphi I_0(2aq)
-c_3^\ast c_2\sin\varphi\cos\varphi I_2(2aq)\Big\}\cos(\Omega_q t)\nonumber\\
&+& i\Big\{ c_3^\ast c_1\sin\varphi \cos^2\varphi I_3(2aq)+
\Big(\big(\vert c_3\vert^2-\vert c_1\vert^2\big)\sin^2\varphi \cos\varphi-c_1^\ast c_3\sin^3\varphi \Big)
I_1(2aq)\Big\}\sin(\Omega_q t)\nonumber\\
&+&\Big\{c_1^\ast c_2\cos^2\varphi I_0(2aq)+c_3^\ast c_2\sin\varphi\cos\varphi I_2(2aq)\Big\}\cos^2(\Omega_q t)
+\Big\{c_2^\ast c_1\cos^2\varphi I_2(2aq)+c_2^\ast c_3\sin\varphi\cos\varphi I_0(2aq)\Big\}\sin^2(\Omega_q t)\Big\}\nonumber\\
&+&\frac{i}{2}\Big\{\Big(\vert c_2\vert^2\cos\varphi -\vert c_1\vert^2\cos^3\varphi
-c_1^\ast c_3\sin\varphi \cos^2\varphi-\vert c_3\vert^2\sin^2\varphi\cos\varphi\Big)I_1(2aq)
-c_3^\ast c_1\sin\varphi \cos^2\varphi I_3(2aq)\Big\}\sin(2\Omega_qt)\Bigg]
\end{eqnarray}

and

\begin{eqnarray}\label{Sig2}
&&\Sigma_{23}(t)
=\frac{2}{C^2} e^{-a^2}\int_0^\infty dq\,q\,e^{-q^2}\Bigg[\Big\{ c_2^\ast c_3\cos^2\varphi I_0(2aq)
-c_2^\ast c_1\sin\varphi\cos\varphi I_2(2aq)\Big\}\cos(\Omega_q t)\nonumber\\
&+& i\Big\{ -c_3^\ast c_1\sin^2\varphi \cos\varphi I_3(2aq)+
\Big(\big(\vert c_3\vert^2-\vert c_1\vert^2\big)\sin\varphi \cos^2\varphi+c_1^\ast c_3\cos^3\varphi \Big)
I_1(2aq)\Big\}\sin(\Omega_q t)\nonumber\\
&+&\Big\{c_2^\ast c_3\sin^2\varphi I_0(2aq)+c_2^\ast c_1\sin\varphi\cos\varphi I_2(2aq)\Big\}\cos^2(\Omega_q t)
+\Big\{c_3^\ast c_2\sin^2\varphi I_2(2aq)+c_1^\ast c_2\sin\varphi\cos\varphi I_0(2aq)\Big\}\sin^2(\Omega_q t)\Big\}\nonumber\\
&+&\frac{i}{2}\Big\{\Big(\vert c_3\vert^2\sin^3\varphi-\vert c_2\vert^2\sin\varphi 
+c_1^\ast c_3\sin^2\varphi \cos\varphi+\vert c_1\vert^2\sin\varphi\cos^2\varphi\Big)I_1(2aq)
+c_3^\ast c_1\sin^2\varphi \cos\varphi I_3(2aq)\Big\}\sin(2\Omega_qt)\Bigg],
\end{eqnarray}
\end{widetext}
where $a=k_0d$, $q=kd$, $\Omega_q=v_Fq/d$, and $I_\nu(x)$ is the $\nu$-th order modified Bessel function of first kind. 
Note that in Eqs. (\ref{Sig1}) and (\ref{Sig2}) there exist two frequencies namely 
$\Omega_q$ and $2\Omega_q$. The former is a result of interference between either FB \& VB or CB \&FB
while the latter arises due to the coupling between CB \& VB.

\subsection{Various types of pseudospin polarization}
We are merely interested in the dynamics of the Gaussian wave packet
of different types of initial pseudo-spin polarization. Here, we will
discuss all the possibilities corresponding to various
combinations of $c_1$, $c_2$, and $c_3$.

\begin{center}
 \textbf{1.~~{\textit{z}}-polarization}
\end{center}
Here, we consider the wave packet was polarized initially along 
$+z$ direction. In this case the possible combinations of $(c_1,c_2,c_3)$
are $(1,0,0)$, $(0,1,0)$, and $(0,0,1)$ corresponding to the eigen states of $S_z$-operator. 
Physically for the choices $(1,0,0)$ and $(0,0,1)$ the initial wave function
was concentrated solely in any of the $rim$ sites while the choice $(0,1,0)$ corresponds to 
the case in which the wave function was entirely located in the $hub$ site initially. 
Let us now discuss all the possibilities one by one. 
 
(i) First we consider $c_1=1$, $c_2=0$, and $c_3=0$ i.e. the initial wave packet
was located in the $rim$ site B only. From Eqs. (\ref{VeloF}), (\ref{Sig1}), and 
(\ref{Sig2}), after a straightforward calculation, we find
$\la v_x(t)\ra=0$ whereas $\la v_y(t)\ra$ takes the following form
\begin{eqnarray}\label{Velc1}
\la v_y(t)\ra= &-&2v_F\cos^2\varphi e^{-a^2}
\int dq qe^{-q^2} I_1 (2aq)
\Big[\cos(2\varphi)\nonumber\\
&\times&\sin(2q\tau)
+2\Big\{1-\cos(2\varphi)\Big\}
\sin(q\tau)\Big],
\end{eqnarray}
where $\tau=v_F t/d$.
Note that Eq. (\ref{Velc1}) is an example of two-frequency ZB for a finite $\alpha$. The
interference between CB and VB leads to the first term in the square bracket in Eq. (\ref{Velc1})
while the second term is appeared as a consequence of the interference between
either CB and FB or VB and FB. When $\alpha=0$ i.e. $\varphi=0$
the second term in the square bracket of Eq. (\ref{Velc1}) contributes nothing to $\la v_y(t)\ra$.
In this case the velocity
average exhibits single frequency ZB determined from the interference between CB and VB which
basically reminiscences the case of graphene. 
In the opposite limit i.e. for $\alpha=1$ we have $\varphi=\pi/4$. In this case, Eq. (\ref{Velc1}) retains only 
the second term in square bracket. Here, $\la v_y(t)\ra$ performs single frequency ZB, determined 
from the coupling between either CB and FB or FB and VB. In other words,
for $\alpha=1$, the interference between CB and VB is completely prevented by FB.

The expectation value of position operator is obtained
by integrating Eq. (\ref{Velc1}) with the initial condition: $\la y(t)\ra=0$ at $t=0$.
Evaluating the $q$-integral in Eq. (\ref{Velc1}) numerically for $a=5$, we show the behavior 
of $\la v_y(t)\ra$ and $\la y(t)\ra$
in Fig. 2 and Fig. 3, respectively for different values of $\alpha$.
\begin{figure}[h!]
\begin{center}\leavevmode
\includegraphics[width=100mm, height=60mm]{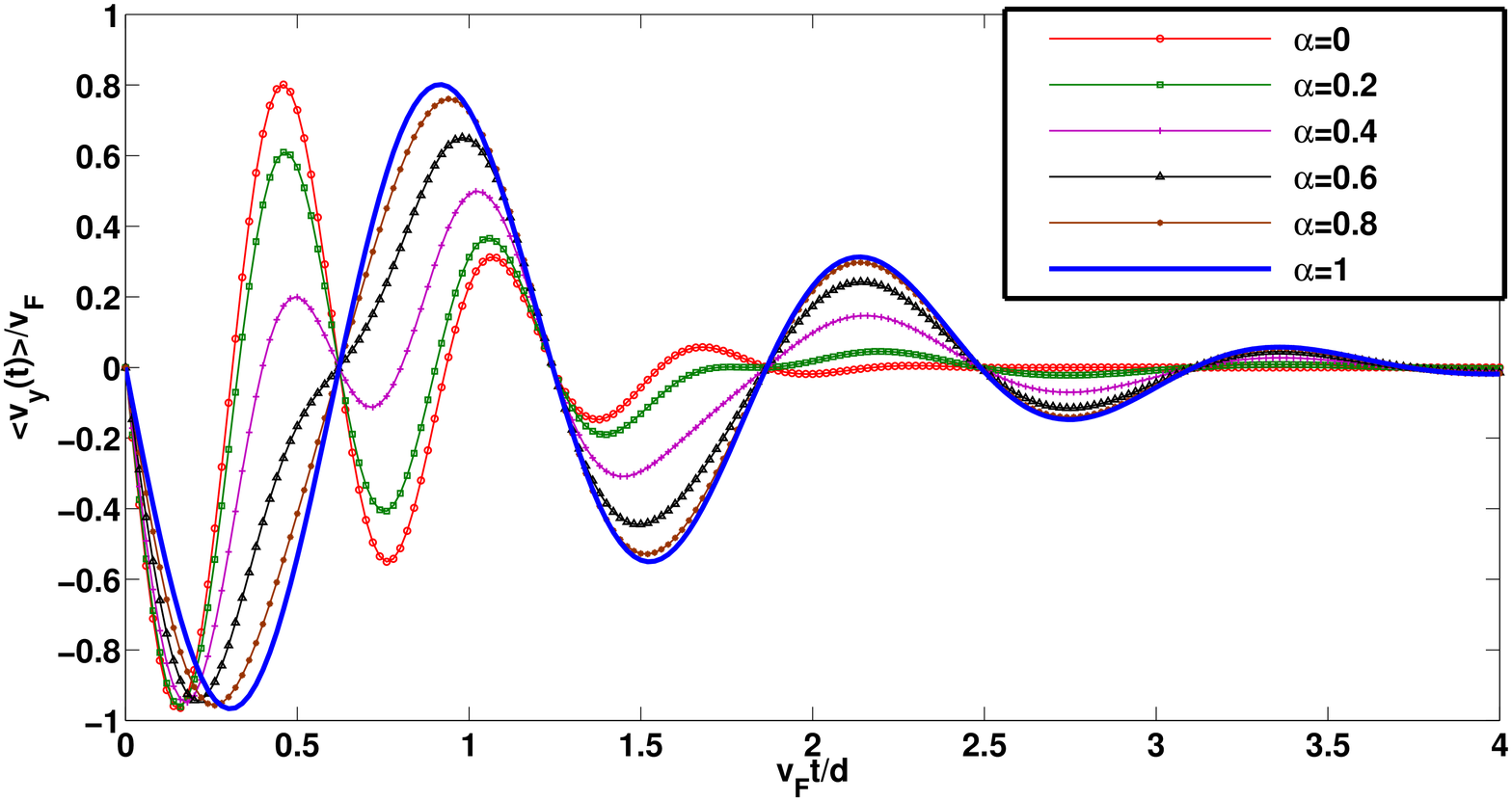}
\caption{Time dependence of the expectation value of velocity operator for $a$=$5$.
We consider initial pseudospin polarization along $z$-direction with 
components $c_1$=$1$, $c_2$=$0$, and $c_3$=$0$. }
\end{center}
\end{figure}

\begin{figure}[h!]
\begin{center}\leavevmode
\includegraphics[width=100mm, height=60mm]{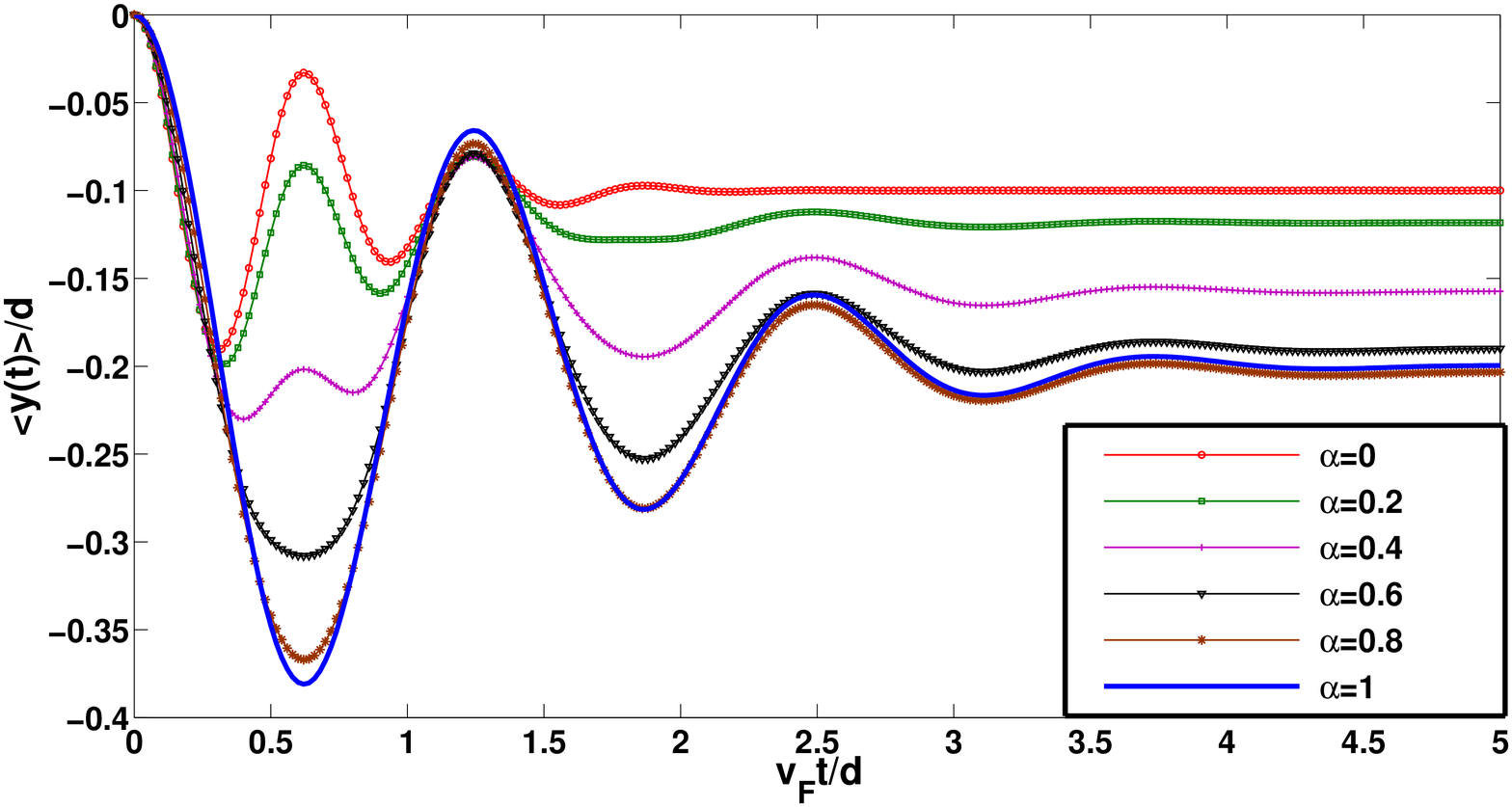}
\caption{Time dependence of the expectation value of position operator for
$a$=$5$. We consider initial pseudospin polarization along $z$-direction with 
components $c_1$=$1$, $c_2$=$0$, and $c_3$=$0$.}
\end{center}
\end{figure}

Both figures clearly demonstrate a crossover from a ZB of frequency $2\Omega_q$ to a ZB of
frequency $\Omega_q$ with the smooth evolution
of $\alpha$ from $0$ to $1$. In other words, as $\alpha$ approaches $1$, the frequency 
of ZB gets reduced to half of the value corresponding to $\alpha=0$. It is difficult to 
comment on the frequency of ZB for $0<\alpha<1$. Due to the interplay of two frequencies, namely,
$2\Omega_q$ and $\Omega_q$ a complicated pattern is obtained for $\alpha=0.4$.
It is worthy to notice that only one frequency either $2\Omega_q$ or $\Omega_q$ roughly dominates
when $\alpha$ is below or above the value $\alpha=0.4$. 
As evident from Fig. 2 and Fig. 3, the resultant ZB is transient in 
character. To understand this transient behavior, it is possible to find an approximate 
analytical expression of Eq. (\ref{Velc1}) when the width of the wave packet is large enough i.e. for
$a\gg1$. In this limit the modified Bessel function $I_1(2aq)$ can be approximated as 
$I_1(2aq)\approx e^{2aq}/\sqrt{4\pi aq}$. With the help of the stationary phase approximation 
we obtain
\begin{eqnarray}\label{Vel_app}
\la v_y(t)\ra \approx &-&\frac{v_F}{2}\cos^2\varphi
\Big[\cos(2\varphi)e^{-\tau^2}\sin(2a\tau)\nonumber\\
&+&2\big\{1-\cos(2\varphi)\big\}e^{-\tau^2/4}\sin(a\tau)\Big].
\end{eqnarray}
Note that the presence of decaying exponential terms in Eq. (\ref{Vel_app}) clearly 
explains the transient behavior of ZB. For $\alpha=0$ the ZB decays rapidly 
due to the presence of the term $e^{-\tau^2}$ in Eq. (\ref{Vel_app}). Here, the characteristic
time scale corresponding to the decay in ZB amplitude is of the order of $d/v_F$.
When $\alpha=1$ ZB decays slowly in comparison to the $\alpha=0$. This behavior is 
justified by the presence of $e^{-\tau^2/4}$ term in Eq. (\ref{Vel_app}). Here the 
characteristic decay time scale is $2d/v_F$. It is also evident from Fig. 2 and 3, 
the second term within the square bracket in Eq. (\ref{Vel_app}) dominates over 
the first term in the intermediate range of $\alpha$ i.e. $0<\alpha<1$.

(ii) Next, we consider another choice, namely, $c_1=0$, $c_2=0$, and $c_3=1$. In this case 
initially the electronic wave function was concentrated only in the other $rim$ site C. Here we find
$\la v_x(t)\ra=0$ and
\begin{eqnarray}\label{Velc2}
\la v_y(t)\ra= &-&2v_F\sin^2\varphi e^{-a^2}
\int dq qe^{-q^2} I_1 (2aq)
\Big[\cos(2\varphi)\nonumber\\
&\times&\sin(2q\tau)
-2\Big\{1+\cos(2\varphi)\Big\}
\sin(q\tau)\Big].
\end{eqnarray}
It is obvious from Eq. (\ref{Velc2}) that $\la v_y(t)\ra=0$ when $\alpha=0$. This 
clearly contradicts the established results corresponding to graphene.
However, the discrepancy arose here is apparent because we are dealing with pseudospin-$1$. For 
graphene the pseudospin component $c_3$ is completely absent. Hence, the results 
corresponding to graphene can not be correctly interpreted from Eq. (\ref{Velc2}).
For a finite $\alpha$, comparing Eq. (\ref{Velc1}) and Eq. (\ref{Velc2}), one may 
say that choices (i) and (ii) impose two-fold differences in ZB. Firstly, the amplitudes differ 
by a $\alpha$-dependent factor, namely, $\sin^2\varphi$ and $\cos^2\varphi$. Secondly, 
the second terms in the square brackets in Eq. (\ref{Velc1}) and (\ref{Velc2}) are 
different which may bring a phase difference in the oscillations. In the 
$a>>1$ limit, we obtain
\begin{eqnarray}
\la v_y(t)\ra \approx &-&\frac{v_F}{2}\sin^2\varphi
\Big[\cos(2\varphi)e^{-\tau^2}\sin(2a\tau)\nonumber\\
&-&2\big\{1+\cos(2\varphi)\big\}e^{-\tau^2/4}\sin(a\tau)\Big].
\end{eqnarray}

(iii) Finally, we consider $c_1=0$, $c_2=1$, and $c_3=0$. The physical meaning
of this particular choice corresponds to the initial localization of the wave function
at the $hub$ site A. In this case, we obtain
$\la v_x(t)\ra=0$ and
\begin{eqnarray}\label{Velc3}
\la v_y(t)\ra= 2v_F\cos(2\varphi)e^{-a^2}
\int dq q e^{-q^2} I_1 (2aq)\sin(2q\tau).
\end{eqnarray}
It is transparent from Eq. (\ref{Velc3}) that ZB consists of only one frequency governed by the energy
difference between CB and VB for all values of $\alpha$. For this specific case, the FB contributes nothing to ZB. 
In the large $a$ limit, we find
\begin{eqnarray}\label{Vel_app3}
\la v_y(t)\ra \approx & &\frac{v_F}{2}\cos(2\varphi)
e^{-\tau^2}\sin(2a\tau).
\end{eqnarray}

For all the choices of $(c_1,c_2,c_3)$, we obtain $\la v_x(t)\ra=0$ but $\la v_y(t)\ra\neq0$. This 
implies that ZB occurs in a direction perpendicular to the 
direction of initial wave vector and pseudospin of the wave packet. Interestingly, we also
note that the behavior of ZB is significantly dependent on the type of the 
initial pseudospin polarization of the wave packet. More specifically, when the initial wave packet
was located entirely in any of the $rim$ sites, as understood from Eq. (\ref{Velc1}) and (\ref{Velc2}),
corresponding ZB consists of two frequencies for $0<\alpha<1$. But when the initial wave packet was in the 
$hub$ site, we obtain single frequency ZB for all $\alpha$ as evident from Eq. (\ref{Velc3}). To
establish all qualitative arguments made earlier we portray the time evolution of
the expectation values of velocity and position operators in Fig. 4 for 
different choices of $(c_1,c_2,c_3)$. For the plots we consider $\alpha=0.5$ and $a=5$.
As illustrated in Fig. 4 the ZBs in position and velocity corresponding to different choices
clearly differ in phase and amplitude. The ZB corresponding to the choice $(c_1,c_2,c_3)=(0,1,0)$
decays more rapidly than the other choices due to the structure of the Eq. (\ref{Vel_app3}). Moreover,
the ZB in position for $(c_1,c_2,c_3)=(1,0,0)$ is negative while the ZBs for other choices are positive. 
\begin{figure}[h!]
\begin{center}\leavevmode
\includegraphics[width=125mm, height=60mm]{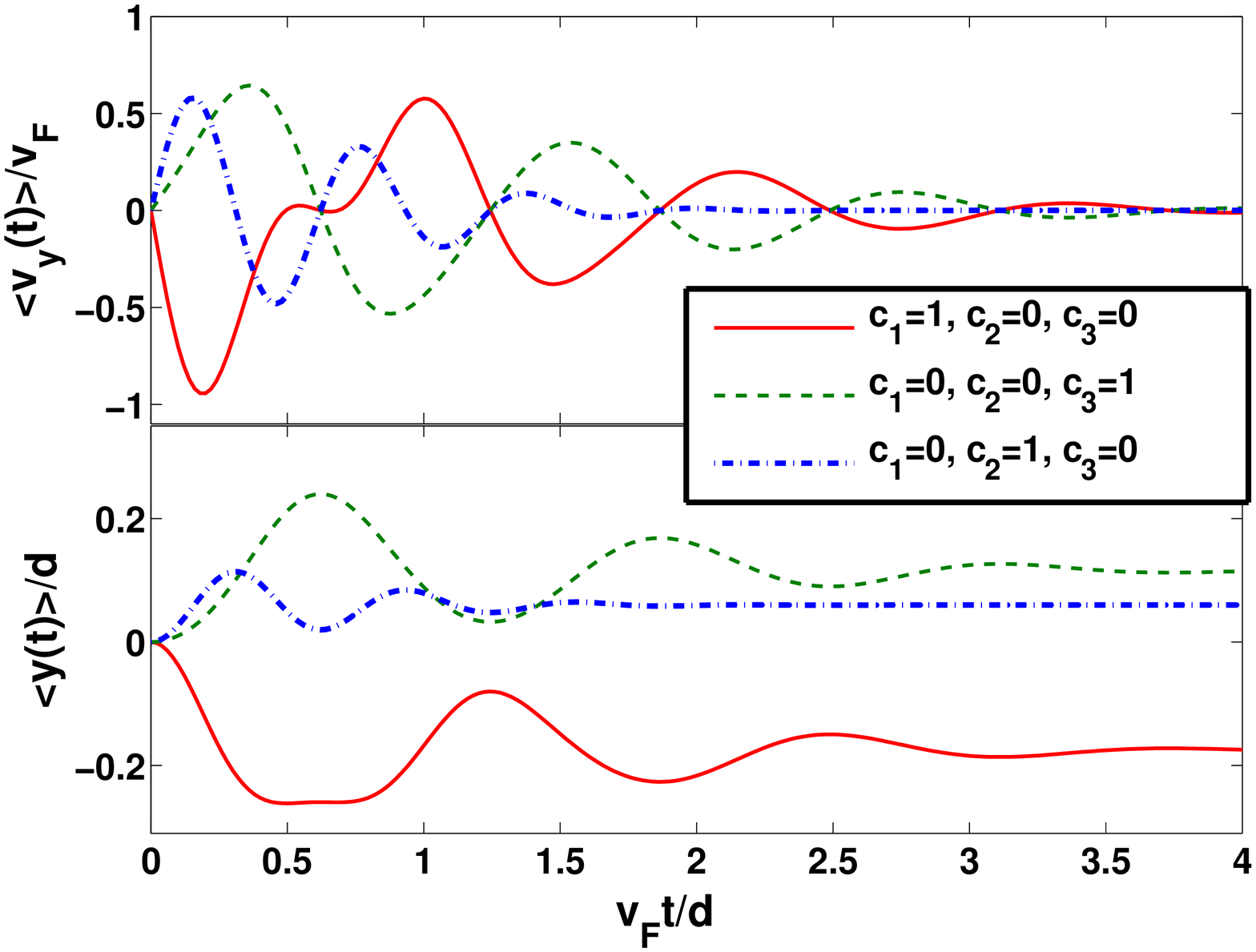}
\caption{Time dependence of the expectation values of velocity and position operator for 
different choices of $(c_1, c_2, c_3)$. Here,
we consider $a=5$ and $\alpha=0.5$.}
\end{center}
\end{figure}

\begin{center}
 \textbf{2.~~{\textit{x}}-polarization}
\end{center}
Now, we seek to study the wave packet dynamics by considering that initial pseudospin polarization
was in the $x$-direction. The operator $S_x$ has three eigenstates corresponding to the eigen values,
namely, $0,\pm1$(in units of $\hbar$). For those states we have following choices of $(c_1, c_2, c_3)$,
namely, $(\sin\varphi, 0, -\cos\varphi)$, $(\cos\varphi, 1, \sin\varphi)$, and $(-\cos\varphi, 1, -\sin\varphi)$.
\begin{figure}[h!]
\begin{center}\leavevmode
\includegraphics[width=105mm, height=60mm]{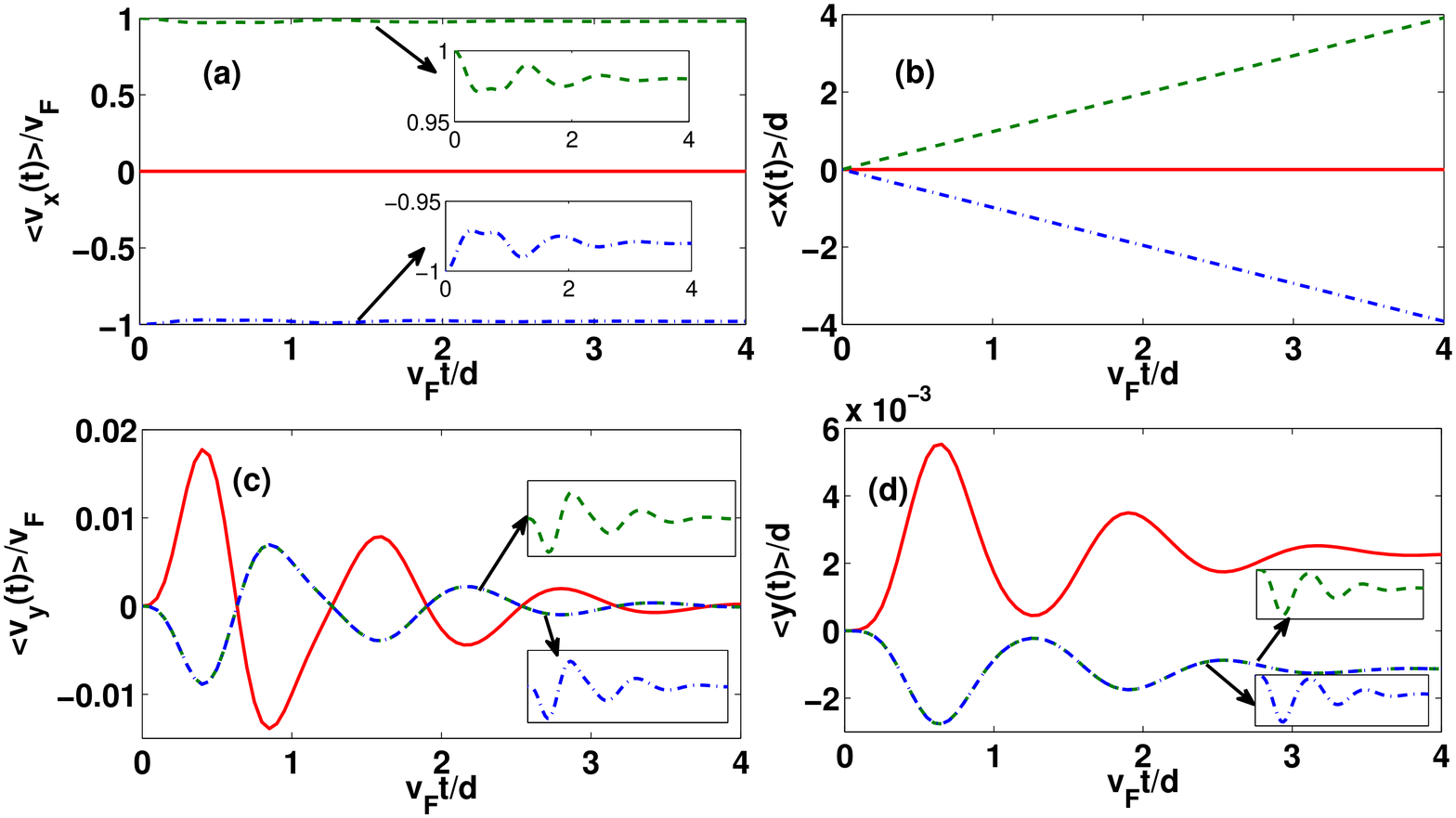}
\caption{Time dependence of the expectation values of velocity and position operator when the 
initial pseudospin was polarized along $x$-direction. Here, solid(red), dashed(green), and 
dot-dashed(blue) lines correspond to different values of $(c_1,c_2,c_3)$, namely, 
$(\sin\varphi, 0, -\cos\varphi)$, $(\cos\varphi, 1, \sin\varphi)$, and $(-\cos\varphi, 1, -\sin\varphi)$,
respectively. We also consider $a=5$ and $\alpha=0.5$.}
\end{center}
\end{figure}
In Fig. 5 the time dependence of the expectation values of position and velocity 
operators are shown corresponding to the above mentioned choices of $(c_1,c_2,c_3)$
for a fixed $\alpha=0.5$. In contrast to the case 1, here both $\la x(t)\ra$ and $\la v_x(t)\ra$ 
do not vanish for all choices of $(c_1,c_2,c_3)$. When $(c_1,c_2,c_3)=(\sin\varphi, 0, -\cos\varphi)$, 
both $\la x(t)\ra$ and $\la v_x(t)\ra$ are zero. Corresponding to the remaining choices of 
$(c_1,c_2,c_3)$, namely, $(\cos\varphi, 1, \sin\varphi)$, and $(-\cos\varphi, 1, -\sin\varphi)$ both
$\la x(t)\ra$ and $\la v_x(t)\ra$ are mirror image of each other individually. Although $\la v_x(t)\ra$
shows a tiny oscillation as depicted in the insets of Fig. 5(a)
but $\la x(t)\ra$ exhibits a linear dependence with time (Fig. 5(b)). On the other hand 
the $y$-component of position and velocity exhibit the expected ZB oscillation. However, their amplitudes are
much smaller in comparison with the case 1. Note that the amplitude of ZB in both $\la x(t)\ra$ and $\la v_x(t)\ra$ for
$(c_1,c_2,c_3)=(\sin\varphi, 0, -\cos\varphi)$ are larger than that corresponding to the other choices. When
$(c_1,c_2,c_3)=(\cos\varphi, 1, \sin\varphi)$ and $(\cos\varphi, 1, -\sin\varphi)$ the corresponding ZBs
coincide with each other.

\begin{center}
 \textbf{3.~~{\textit{y}}-polarization}
\end{center}
Finally, we consider the pseudospin associated with the initial wave packet was polarized along
$y$-direction. Like $S_x$ the operator $S_y$ has also eigen values, namely, $0,\pm1$ (in units of $\hbar$).
In this situation the options of $(c_1, c_2, c_3)$ are
$(\sin\varphi, 0, \cos\varphi)$, $(-i\cos\varphi, 1, i\sin\varphi)$, and $(i\cos\varphi, 1, -i\sin\varphi)$.
Fig. 6 shows the behavior of ZB in position and velocity corresponding to those choices of $(c_1,c_2,c_3)$.
For this particular case, ZB appears only in $y$-direction. Different possibilities of $(c_1,c_2,c_3)$ introduce 
a phase in the oscillation.
\begin{figure}[h!]
\begin{center}\leavevmode
\includegraphics[width=170mm, height=80mm]{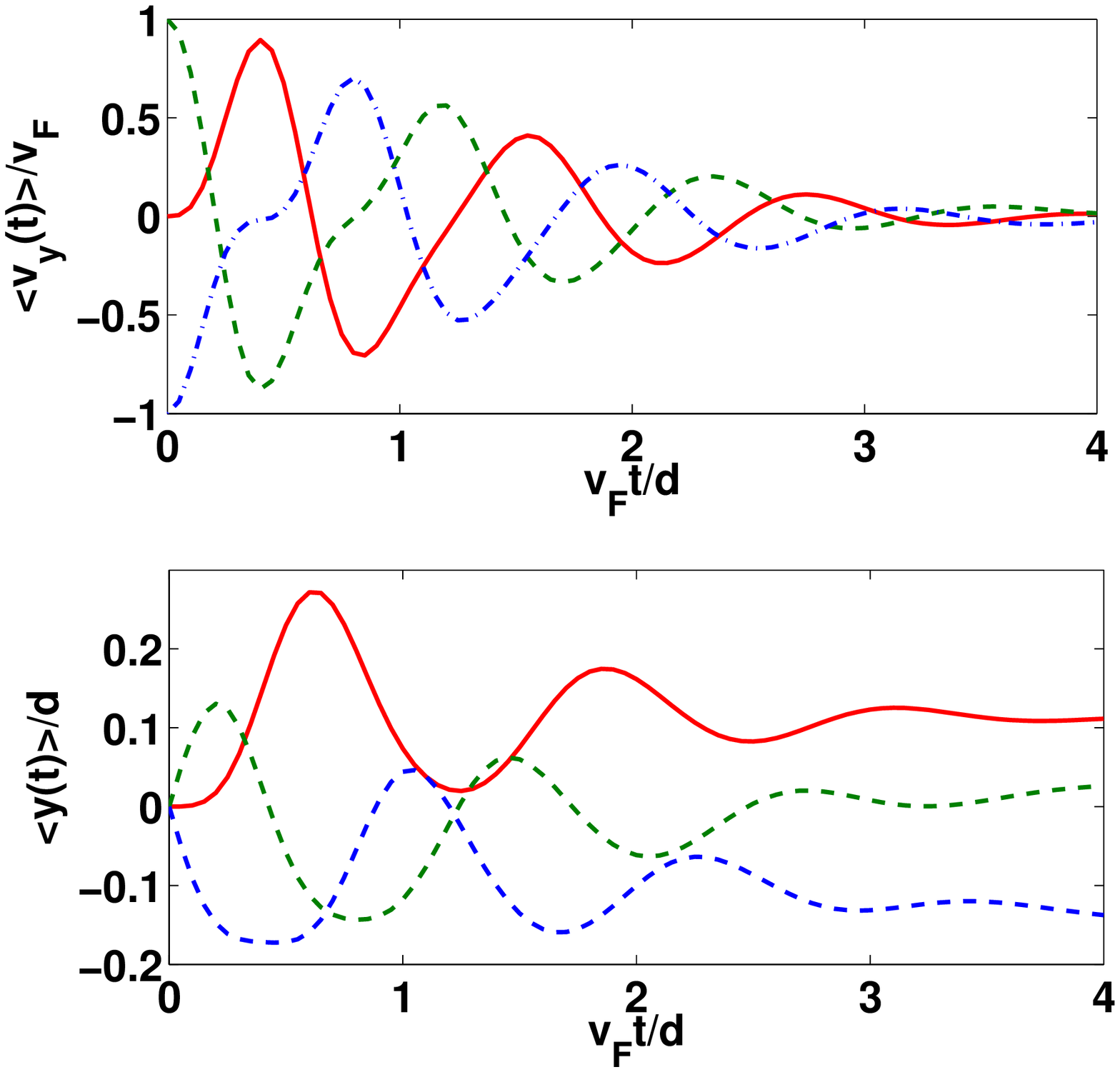}
\caption{Time dependence of the expectation values of velocity and position operator when the 
initial pseudospin was polarized along $y$-direction. Here, solid(red), dashed(green), and 
dot-dashed(blue) lines correspond to different values of $(c_1,c_2,c_3)$, namely, 
$(\sin\varphi, 0, \cos\varphi)$, $(-i\cos\varphi, 1, i\sin\varphi)$, and $(i\cos\varphi, 1, -i\sin\varphi)$,
respectively. We also consider $a=5$ and $\alpha=0.5$.}
\end{center}
\end{figure}

\section{In presence of a magnetic field}

\subsection{Energy spectrum}
As a consequence of an external transverse magnetic field ${\bs{\mathcal{B}}}=\mathcal{B}\hat{z}$, the continuous
energy spectrum of the conic band is redistributed in the form of following Landau levels :
\begin{eqnarray}\label{Energy}
\varepsilon_{n,\zeta}^\lambda=\lambda \, \gamma_B\sqrt{n+\chi_\zeta},
\end{eqnarray}
where
$\lambda=\pm1$ denotes either CB or VB, $n=0,1,2,...$ is the LL index,
$\gamma_B=\sqrt{2}\hbar v_F/l_0$ with $l_0=\sqrt{\hbar/(e\mathcal{B})}$ 
being the magnetic length and $\chi_\zeta=[1-\zeta\cos(2\varphi)]/2$.
It is important to note that the magnetic field can not alter the fate of the FB. It still 
retains its zero energy states.

Choosing the vector potential $\bs{\mathcal{A}}$ corresponding to ${\bs{\mathcal{B}}}$ in the
Landau gauge as ${\bs{\mathcal{A}}}=(-\mathcal{B}y,0,0)$,
we find the conic band eigenfunctions corresponding to the K-valley as\cite{dice_MagTr}
\begin{eqnarray} 
\psi_{n,k_x}^{\lambda, {\rm K}}({\bf r})=\frac{1}{\sqrt{2}}\left(
\begin{array}{c}
\frac{\sqrt{n(1-\chi_+)}}{\sqrt{n+\chi_+}} \Phi_{n-1}(y)\\
\lambda \Phi_n(y)\\
\frac{\sqrt{(n+1)\chi_+}}{\sqrt{n+\chi_+}}\Phi_{n+1}(y)
\end{array}\right)\frac{e^{ik_xx}}{\sqrt{2\pi}}
\end{eqnarray}

and
\begin{eqnarray}
\psi_{0,k_x}^{\lambda,{\rm K}}({\bf r})=\frac{1}{\sqrt{2}}\left(
\begin{array}{c}
0\\
\lambda\Phi_0(y)\\
\Phi_1(y)
\end{array}\right)\frac{e^{ik_xx}}{\sqrt{2\pi}}
\end{eqnarray}
for $n>0$ and $n=0$, respectively.
Here, $\Phi_n(y)=\sqrt{1/(2^nn!\sqrt{\pi}l_0)} e^{-(y-y_0)^2/(2l_0^2)}
H_n[(y-y_0)/l_0]$, with $y_0=l_0^2k_x$, is the standard harmonic oscillator wave function.

On the other hand, despite its zero energy the FB wave functions for K-valley are obtained as\cite{dice_MagTr}
\begin{eqnarray}
\psi_{n,k_x}^{{\rm F}, {\rm K}}({\bf r})=\left(
\begin{array}{c}
-\frac{\sqrt{(n+1)\chi_+}}{\sqrt{n+\chi_+}}\Phi_{n-1}(y)\\
0\\
\frac{\sqrt{n(1-\chi_+)}}{\sqrt{n+\chi_+}}\Phi_{n+1}(y)
\end{array}\right)\frac{e^{ik_xx}}{\sqrt{2\pi}}
\end{eqnarray}
and
\begin{eqnarray}
\psi_{0,k_x}^{{\rm F},{\rm K}}({\bf r})=\left(
\begin{array}{c}
0\\
0\\
\Phi_0(y)
\end{array}\right)\frac{e^{ik_xx}}{\sqrt{2\pi}},
\end{eqnarray}
for $n>0$ and $n=0$, respectively. Note that the 
states in the FB is infinitely degenerate.

\subsection{Time evolution}
Here we attempt to study the cyclotron dynamics of a quasiparticle represented
by a wave packet. It should be mentioned here that we have considered the momentum-space
wave packet in the case of zero magnetic field, but here we prefer to adopt position-space 
wave packet to circumvent calculation difficulties.

The initial Gaussian wave packet is chosen as
\begin{eqnarray}\label{timM1}
\Psi({\bf r},0)=\frac{1}{\sqrt{\pi}l_0C} \exp\Big(-\frac{r^2}{2l_0^2}+i\frac{p_{0x}x}{\hbar}\Big)\left(
\begin{array}{c}
c_1\\
c_2\\
c_3
\end{array}\right),
\end{eqnarray}
where $p_{0x}$ is the initial momentum and the constants $c_1$, $c_2$, $c_3$ play the same 
role as mentioned in the case of zero external field.

To find the wave packet at a later time $t$, we need to construct suitable propagator that 
can describe the desired time evolution. In this way we follow the Green's function technique 
given in Ref. [\onlinecite{zb2d3}]
with appropriate modifications. 

The time evolved wave packet can be found as 
\begin{eqnarray}\label{timM2}
\Psi({\bf r},t)=\int d{\bf r^\prime}G({\bf r},{\bf r^\prime},t)\Psi({\bf r},0),
\end{eqnarray}
where the propagator or the Green's function is given by
\begin{eqnarray}\label{HamK}
G({\bf r},{\bf r^\prime},t)=
\begin{pmatrix}
    G_{11}  & G_{12}  & G_{13} \\
    G_{21}  & G_{22}  & G_{23} \\
    G_{31}  & G_{32}  & G_{33}
\end{pmatrix}.
\end{eqnarray}
The matrix elements of $G({\bf r},{\bf r^\prime},t)$ are defined as
\begin{eqnarray}
G_{lm}({\bf r},{\bf r^\prime},t)=\sum_{\lambda=\pm}\sum_{n=0}^\infty
\psi_{n,k_x,l}^\lambda({\bf r}, t)\psi_{n,k_x,m}^{\lambda^\ast}({\bf r}, 0),
\end{eqnarray}
where the time evolved state is given by
$\psi_{n,k_x,l}^\lambda({\bf r}, t)=\psi_{n,k_x,l}^{\lambda}({\bf r}, 0)e^{-i\varepsilon_n^\lambda t/\hbar}$. 

The corresponding matrix elements i.e. $G_{lm}$'s can be found from 
the following equation
\begin{widetext}
\begin{eqnarray}\label{timM3}
 \begin{pmatrix}
  G_{11}\\
  G_{21}\\
  G_{31}\\
  G_{12}\\
  G_{22}\\
  G_{32}\\
  G_{13}\\
  G_{23}\\
  G_{33}
 \end{pmatrix}
=\frac{1}{2\pi}\int dk_x e^{ik_x(x-x^\prime)}\sum_{n=0}^\infty
 \begin{pmatrix}
 f_{n+1}(t)\phi_n(y-y_c)\phi_n(y^\prime-y_c)\\
 g_{n+1}(t)\phi_{n+1}(y-y_c)\phi_n(y^\prime-y_c)\\
 h_{n+1}(t)\phi_{n+2}(y-y_c)\phi_n(y^\prime-y_c)\\
 g_{n+1}(t)\phi_n(y-y_c)\phi_{n+1}(y^\prime-y_c)\\
 q_n(t)\phi_n(y-y_c)\phi_n(y^\prime-y_c)\\
 p_n(t)\phi_{n+1}(y-y_c)\phi_n(y^\prime-y_c)\\
 h_{n+1}(t)\phi_n(y-y_c)\phi_{n+2}(y^\prime-y_c)\\
 p_n(t)\phi_n(y-y_c)\phi_{n+1}(y^\prime-y_c)\\
 r_n(t)\phi_n(y-y_c)\phi_n(y^\prime-y_c)
 \end{pmatrix},
\end{eqnarray}
where $f_n(t)=A_n\cos(\delta_nt)+B_n$, $g_n(t)=-iC_n\sin(\delta_nt)$, $h_n(t)=D_n(\cos(\delta_nt)-1)$,
$q_n(t)=\cos(\delta_nt)$, $p_n(t)=-iF_n\sin(\delta_nt)$, and $r_n(t)=B_{n-1}\cos(\delta_{n-1}t)+A_{n-1}$ with
$A_n=n(1-\chi_+)/(n+\chi_+)$, $B_n=(n+1)\chi_+/(n+\chi_+)$, $C_n=\sqrt{n(1-\chi_+)/(n+\chi_+)}$,
$F_n=\sqrt{(n+1)\chi_+/(n+\chi_+)}$, and $\delta_n=\gamma_B\sqrt{n+\chi_+}/\hbar$.\\

%
%

Now using Eq. (\ref{timM1}), (\ref{timM2}), and (\ref{timM3}), we find 
the wave packet at a later time $t$ as
\begin{eqnarray}
\left(
\begin{array}{c}
\Psi_1 ({\bf r},t)\\
\Psi_2({\bf r},t)\\
\Psi_3({\bf r},t)
\end{array}\right)
=\frac{1}{\sqrt{2}\pi l_0}\sum_{n=0}^\infty \frac{1}{2^n n!}
\int du e^{\Gamma(x,y,u)}(-u)^n
\left(
\begin{array}{c}
\Sigma_1 (t,u)\\
\Sigma_2(t,u)\\
\Sigma_3(t,u)
\end{array}\right),
\end{eqnarray}

where
$\Gamma(x,y,u)=iux/l_0-(p_{0x}l_0/\hbar-u)^2/2-u^2/4-(y/l_0-u)^2/2$ 
and $\Sigma_\iota(t,u)$'s with $\iota=1,2,3$ are given by the following matrix equation

\begin{eqnarray}\label{Ham}
 \begin{pmatrix}
\Sigma_1(t,u)\\
\Sigma_2(t,u)\\
\Sigma_3(t,u)
\end{pmatrix}
=\frac{1}{C}
\begin{pmatrix}
f_{n+1}(t) H_n(u)  & -\frac{1}{\sqrt{2}}\frac{g_{n+1}(t)}{\sqrt{n+1}}uH_n(u) &  
\frac{1}{2}\frac{h_{n+1}(t)}{\sqrt{(n+1)(n+2)}}u^2H_n(u)  \\
\frac{1}{\sqrt{2}}\frac{g_{n+1}(t)}{\sqrt{n+1}} H_{n+1}(u)  & q_n(t)H_n(u)  & 
-\frac{1}{\sqrt{2}}\frac{p_n(t)}{\sqrt{n+1}}uH_n(u) \\
\frac{1}{2}\frac{h_{n+1}(t)}{\sqrt{(n+1)(n+2)}} H_{n+2}(u) & \frac{1}{\sqrt{2}}\frac{p_n(t)}{\sqrt{n+1}}H_{n+1}(u) &
 r_n(t)H_n(u)
\end{pmatrix}
\begin{pmatrix}
 c_1\\
 c_2\\
 c_3
\end{pmatrix}.
\end{eqnarray}

%
%

\end{widetext}

\subsection{Expectation values}
The expectation value of the velocity operator can be written in the following form as
\begin{eqnarray}\label{VelM}
 \begin{pmatrix}
 \la v_x(t)\ra\\
\la v_y(t)\ra
 \end{pmatrix}
 =2v_F
 \begin{pmatrix}
  {\rm Re}\\
  {\rm Im}
 \end{pmatrix}
\Big[\cos\varphi I_{12}(t)+\sin\varphi I_{23}(t)\Big],
\end{eqnarray}
where,
$I_{\mu\sigma}(t)$ 
is defined as 
$I_{\mu\sigma}(t)=\int\Psi_\mu^\ast({\bf r},t)\Psi_\sigma({\bf r},t)dxdy$.
After a straightforward calculation
we evaluate $I_{12}(t)$ and $I_{23}(t)$ as
\begin{widetext}
\begin{eqnarray}\label{I1}
 I_{12}(t)&=&\sqrt{\frac{2}{3}}\frac{1}{C^2}\sum_{n=0}^\infty \frac{1}{2^n n!}\Bigg[c_1^\ast c_2 f_{n+1}^\ast q_n\xi_{2n}
 +\frac{1}{\sqrt{2(n+1)}}\Big\{\vert c_1\vert^2 f_{n+2}^\ast g_{n+1}+c_1^\ast c_3 f_{n+1}^\ast p_n
 +\vert c_2\vert^2 g_{n+1}^\ast q_n\Big\}\xi_{2n+1}\nonumber\\
 &+&\frac{1}{2\sqrt{(n+1)(n+2)}}\Big\{ c_3^\ast c_2 h_{n+1}^\ast q_n+c_2^\ast c_1 g_{n+2}^\ast g_{n+1}
 +c_2^\ast c_3\sqrt{\frac{n+2}{n+1}} g_{n+1}^\ast p_n\Big\}\xi_{2n+2}\nonumber\\
 &+&\frac{1}{2\sqrt{2(n+1)(n+2)(n+3)}}\Big\{ c_3^\ast c_1 h_{n+2}^\ast g_{n+1}+
 \vert c_3\vert^2\sqrt{\frac{n+3}{n+1}} h_{n+1}^\ast p_n\Big\}\xi_{2n+3}\Bigg]
\end{eqnarray}

and
 
\begin{eqnarray}\label{I2}
 I_{23}(t)&=&\sqrt{\frac{2}{3}}\frac{1}{C^2}\sum_{n=0}^\infty \frac{1}{2^n n!}\Bigg[\Big\{c_1^\ast c_2 g_{n+1}^\ast p_n
 +c_2^\ast c_3 q_n^\ast r_n\Big\}\xi_{2n}
 +\frac{1}{\sqrt{2(n+1)}}\Big\{\vert c_1\vert^2 g_{n+2}^\ast h_{n+1}+c_1^\ast c_3 g_{n+1}^\ast r_{n+1}\nonumber\\
 &+&\vert c_2\vert^2 q_{n+1}^\ast p_n
 +\vert c_3\vert^2 p_n^\ast r_n\Big\}\xi_{2n+1}
 +\frac{1}{2\sqrt{(n+1)(n+2)}}\Big\{ c_2^\ast c_1 q_{n+2}^\ast h_{n+1}+c_3^\ast c_2 p_{n+1}^\ast p_n\Big\}\xi_{2n+2}\nonumber\\
 &+&\frac{1}{2\sqrt{2(n+1)(n+2)(n+3)}} c_3^\ast c_1\frac{p_{n+2}^\ast h_{n+1}}{\sqrt{(n+1)(n+2)(n+3)}}\xi_{2n+3}\Bigg],
\end{eqnarray}
where
\begin{eqnarray}
\xi_m=(-1)^m\Big(\frac{2}{3}\Big)^{\frac{m}{2}}{\rm exp}\bigg(-\frac{p_{0x}^2l_0^2}{3\hbar^2}\bigg)
\frac{H_m(i\sqrt{\frac{2}{3}}\frac{p_{0x}l_0}{\hbar})}{(2i)^m}.
\end{eqnarray}

\end{widetext}

\subsection{Different choices of initial pseudospin polarization}
Likewise the zero magnetic field case we discuss here the behavior of ZB
for various choices of initial pseudospin polarization corresponding to the 
different values of $c_1$, $c_2$, and $c_3$.

\begin{center}
 \textbf{1.~~{\textit{z}}-polarization}
\end{center}
We consider the pseudospin associated with the initial wave packet is 
polarized along the $z$-direction. Similar to the zero field case, here we also have 
three distinct choices of $(c_1,c_2,c_3)$, namely, $(1,0,0)$, $(0,1,0)$, and 
$(0,0,1)$. In the following we demonstrate how different choices of $(c_1,c_2,c_3)$
lead to modify the structure of ZB.

(i) For $(c_1,c_2,c_3)=(1,0,0)$ we find from Eq. (\ref{I1}) and Eq. (\ref{I2})
\begin{eqnarray}
I_{12}(t)=\frac{1}{\sqrt{3}}\sum_{n=0}^\infty \frac{f_{n+2}^\ast g_{n+1}}{2^n n!\sqrt{n+1}}
\xi_{2n+1}.
\end{eqnarray}
\begin{eqnarray}
I_{23}(t)=\frac{1}{\sqrt{3}}\sum_{n=0}^\infty \frac{g_{n+2}^\ast h_{n+1}}{2^n n!\sqrt{n+1}}
\xi_{2n+1}.
\end{eqnarray}
Note that the products $f_{n+2}^\ast g_{n+1}$ and $g_{n+2}^\ast h_{n+1}$ are purely imaginary.
As a result, we readily obtain from Eq. (\ref{VelM}) that $\la v_x(t)\ra=0$ and 
\begin{eqnarray}\label{Velz1}
\la v_y(t)\ra&=&\frac{2v_F}{\sqrt{3}}\sum_{n=0}^\infty
\frac{\xi_{2n+1}}{2^n n!\sqrt{n+1}}{\rm Im}
\Big(\cos\varphi\, f_{n+2}^\ast g_{n+1}\nonumber\\
&+&\sin\varphi\, g_{n+2}^\ast h_{n+1}\Big).
\end{eqnarray}

(ii) For $(c_1,c_2,c_3)=(0,1,0)$ we find $\la v_x(t)\ra=0$ and 
\begin{eqnarray}\label{Velz2}
\la v_y(t)\ra&=&\frac{2v_F}{\sqrt{3}}\sum_{n=0}^\infty
\frac{\xi_{2n+1}}{2^n n!\sqrt{n+1}}{\rm Im}
\Big(\cos\varphi\, g_{n+1}^\ast q_{n}\nonumber\\
&+&\sin\varphi\, q_{n+1}^\ast p_{n}\Big).
\end{eqnarray}

(iii) When $(c_1,c_2,c_3)=(0,0,1)$, it is obtained that $\la v_x(t)\ra=0$ and 
\begin{eqnarray}\label{Velz3}
\la v_y(t)\ra&=&\frac{2v_F}{\sqrt{3}}\sum_{n=0}^\infty
\frac{1}{2^n n!\sqrt{n+1}}{\rm Im}
\Big(\sin\varphi \, p_{n}^\ast r_{n}\xi_{2n+1}\nonumber\\
&+&\frac{\cos\varphi}{2\sqrt{(n+1)(n+2)}} h_{n+1}^\ast p_{n}\xi_{2n+3}\Big).
\end{eqnarray}
A careful inspection of Eq. (\ref{Velz1})-(\ref{Velz3}) reveals that the structure
of the ZB appeared in velocity significantly depends on different choices of 
$(c_1,c_2,c_3)$.
\begin{figure}[h!]
\begin{center}\leavevmode
\includegraphics[width=105mm, height=50mm]{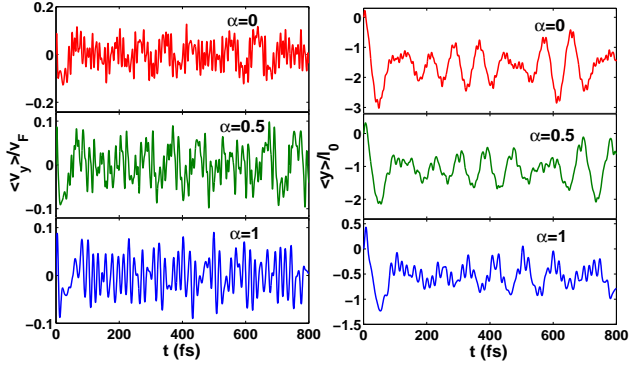}
\caption{Expectation values of position and velocity operators in presence of a 
transverse magnetic field for pseudo spin polarization along $z$-direction with 
components $(c_1,c_2,c_3)$=$(1,0,0)$.
Here, we consider $k_{0x}=2\times 10^8$ m$^{-1}$ and $l_0$=$8.11$ nm.}
\end{center}
\end{figure}
In all cases the velocity average undergoes multi-frequency transverse ZB 
governed by the interference among different Landau levels. 
In Fig. 7 - Fig. 9 we portray the time dependence of the expectation values of 
velocity and position operators. With appropriate initial conditions taken into 
account corresponding position expectation values can be 
obtained by integrating Eqs. (\ref{Velz1})-(\ref{Velz3}).
The actual initial condition is $\la y\ra=\la y_0\ra=k_{0x}l_0^2$
at $t=0$. To calculate $\la y(t)\ra$ we choose
the following initial condition : $\la y\ra=0$ at $t=0$. The expectation values 
of position operator as illustrated in Fig. 7 - Fig. 9 differ from their actual
values at most by a constant shift $\la y_0\ra$. 
We perform all the calculations for a constant magnetic field $\mathcal{B}=10$ T for 
which the magnetic length scale becomes $l_0=8.11$ nm. We also consider 
the width of the wave packet as $d=l_0$. Fig. 7 illustrates the ZB appeared 
in velocity and position for various values of $\alpha$. In this case we 
choose $(c_1,c_2,c_3)=(1,0,0)$ and $k_{0x}=2\times 10^8$ m$^{-1}$. It is clear 
from Fig. 7, the ZBs appeared in both position and velocity undergo perform 
permanent oscillations and oscillatory patterns depend significantly on $\alpha$.

\begin{figure}[h!]
\begin{center}\leavevmode
\includegraphics[width=105mm, height=50mm]{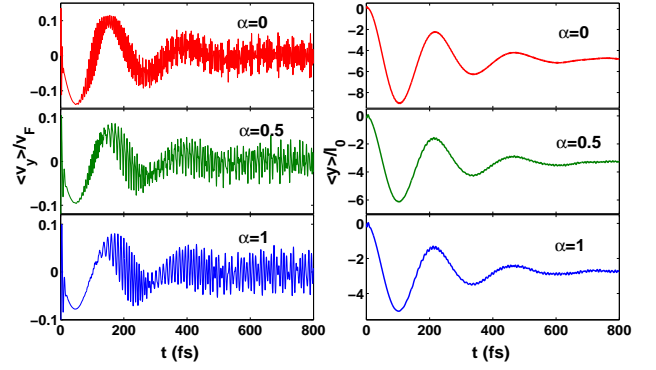}
\caption{Expectation values of position and velocity operators in presence of a 
transverse magnetic field for pseudo spin polarization along $z$-direction with 
components $(c_1,c_2,c_3)=(1,0,0)$.
Here, we consider $k_{0x}$=$6\times 10^8$ m$^{-1}$ and $l_0$=$8.11$ nm.}
\end{center}
\end{figure}

\begin{figure}[h!]
\begin{center}\leavevmode
\includegraphics[width=105mm, height=50mm]{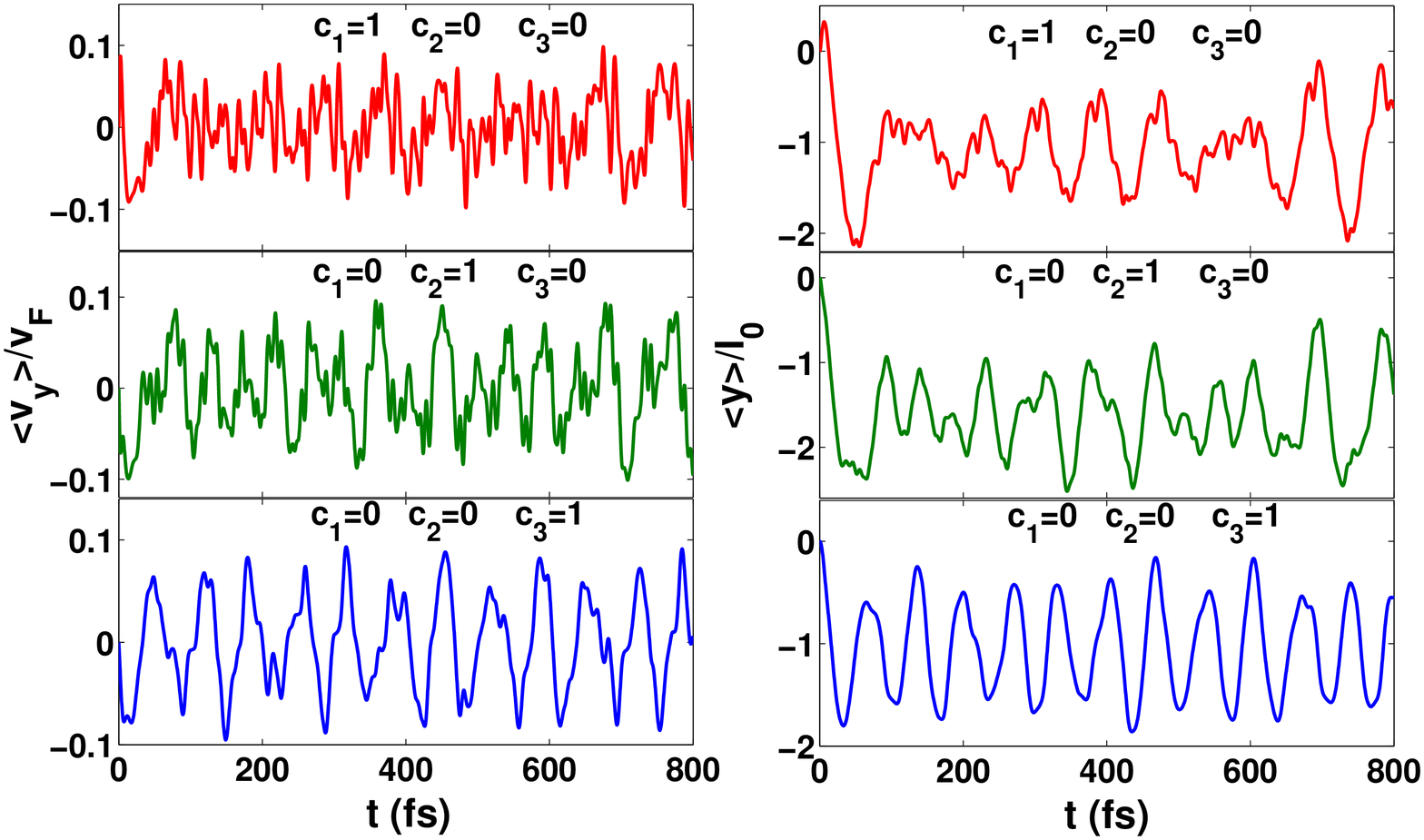}
\caption{Expectation values of position and velocity operators in presence of  a 
transverse magnetic field for pseudospin polarization along $z$-direction with 
different combinations of $(c_1, c_2, c_3)$.
Here, we consider $k_{0x}$=$2\times 10^8$ m$^{-1}$, $l_0$=$8.11$ nm, and $\alpha$=$0.5$.}
\end{center}
\end{figure}

To plot Fig. 8 we repeat the calculations for a higher value of $k_{0x}$, namely,
$k_{0x}=6\times 10^8$ m$^{-1}$. In this case the ZB in position exhibits transient 
character. However, for ZB in velocity a highly oscillatory pattern is superimposed on the transient character.
Note that the locations of maxima and minima are almost insensitive to $\alpha$. Only the
amplitude of ZB changes with $\alpha$.

Fig. 9 describe the behavior of ZB corresponding to different possibilities of $(c_1,c_2,c_3)$.
For the plots, we have taken $k_{0x}=2\times 10^8$ m$^{-1}$ and $\alpha=0.5$. Here, the oscillatory
pattern is significantly different for different initial pseudospin polarization.

\begin{center}
 \textbf{2.~~{\textit{x}}-polarization}
\end{center}
Here, we consider that the initial wave packet was polarized along $x$-direction and the 
resultant behaviors are shown in Fig. 10 and Fig. 11. As mentioned
earlier we have following choices of $(c_1,c_2,c_3)$, namely, $(\sin\varphi,0,-\cos\varphi)$, 
$(\cos\varphi,1,\sin\varphi)$, and $(-\cos\varphi,1,-\sin\varphi)$. In this case the values of parameters
are taken as $k_{0x}$=$2\times10^8$ m$^{-1}$, $\alpha$=$0.5$, and $l_0$=$8.11$ nm. 

\begin{figure}[h!]
\begin{center}\leavevmode
\includegraphics[width=155mm, height=70mm]{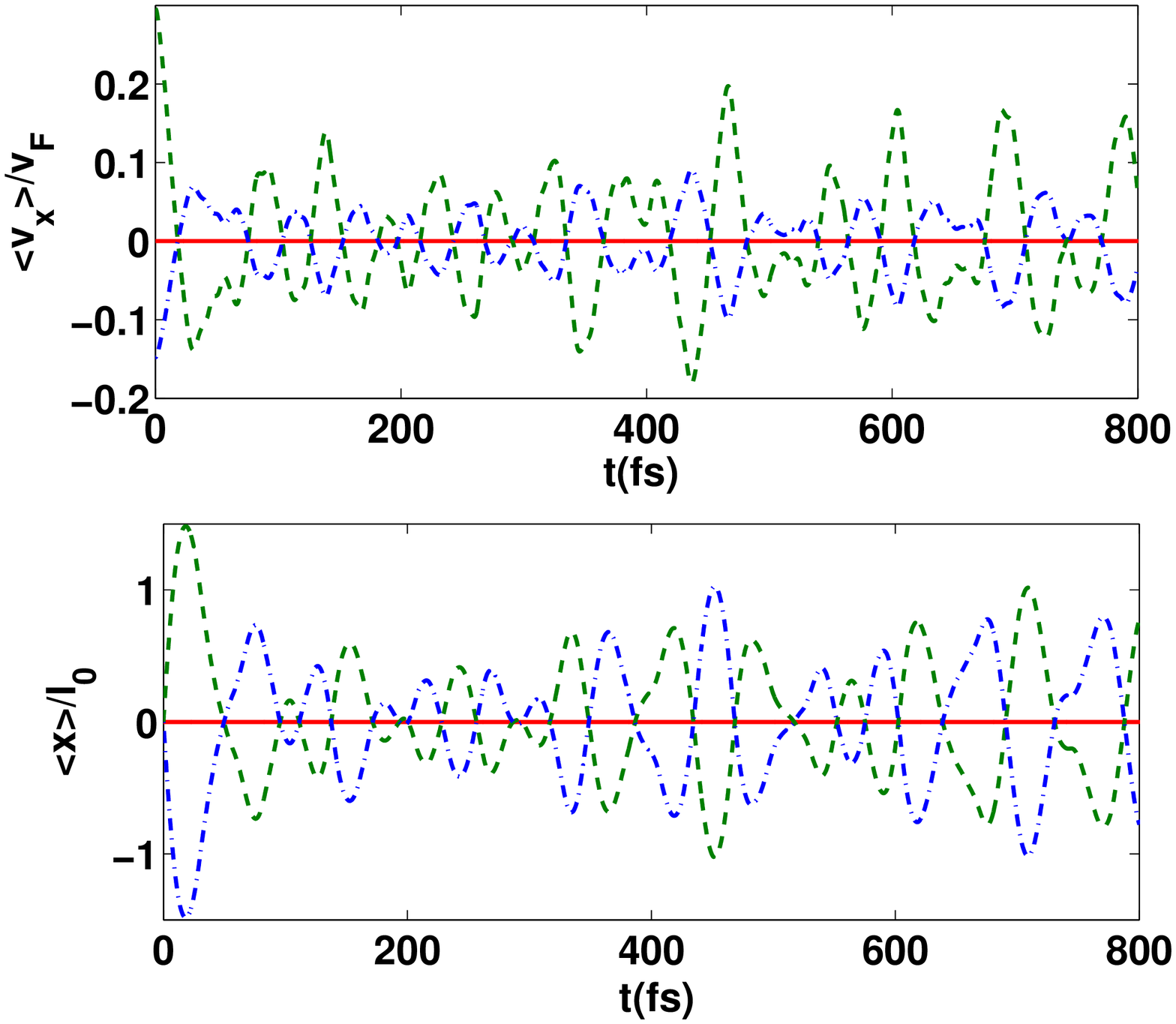}
\caption{Time dependence of the expectation values of $x$-component of velocity and position operators when the 
initial pseudospin was polarized along $x$-direction. Here, solid(red), dashed(green), and 
dot-dashed(blue) lines correspond to different values of $(c_1,c_2,c_3)$, namely, 
$(\sin\varphi, 0, -\cos\varphi)$, $(\cos\varphi, 1, \sin\varphi)$, and $(-\cos\varphi, 1, -\sin\varphi)$,
respectively. We also consider $k_{0x}$=$2\times 10^8$ m$^{-1}$, $l_0$=$8.11$ nm, and $\alpha$=$0.5$.}
\end{center}
\end{figure}

Interestingly we find that the $x$-component 
of the expectation
values of velocity and position operators are non-zero as evident from Fig. 10.
However, for $(c_1,c_2,c_3)=(\sin\varphi,0,-\cos\varphi)$,
it is obtained that $\la v_x(t)\ra=0$ and $\la x(t)\ra=0$. The expectation values of $x$ corresponding to other choices
of $(c_1,c_2,c_3)$ are mirror images of each other. This particular feature was also reflected in the case of 
zero magnetic field. However, the values of $\la v_x(t)\ra$ corresponding to $(c_1,c_2,c_3)=(\cos\varphi,1,\sin\varphi)$
and $(c_1,c_2,c_3)=(-\cos\varphi,1,-\sin\varphi)$ are not exactly mirror images of each other. In fact, the amplitude of 
$\la v_x(t)\ra$ in the first case is greater than that in the second case.

On the other hand, the time dependence of the expectation values of $y$ and $v_y$ are portrayed in Fig. 11.
Both $\la y(t)\ra$ and $\la v_y(t)\ra$ exhibit regular oscillations for $(c_1,c_2,c_3)=(\sin\varphi,0,-\cos\varphi)$.
Irregularities in the oscillations appear for other choices of $(c_1,c_2,c_3)$.

\begin{figure}[h!]
\begin{center}\leavevmode
\includegraphics[width=130mm, height=65mm]{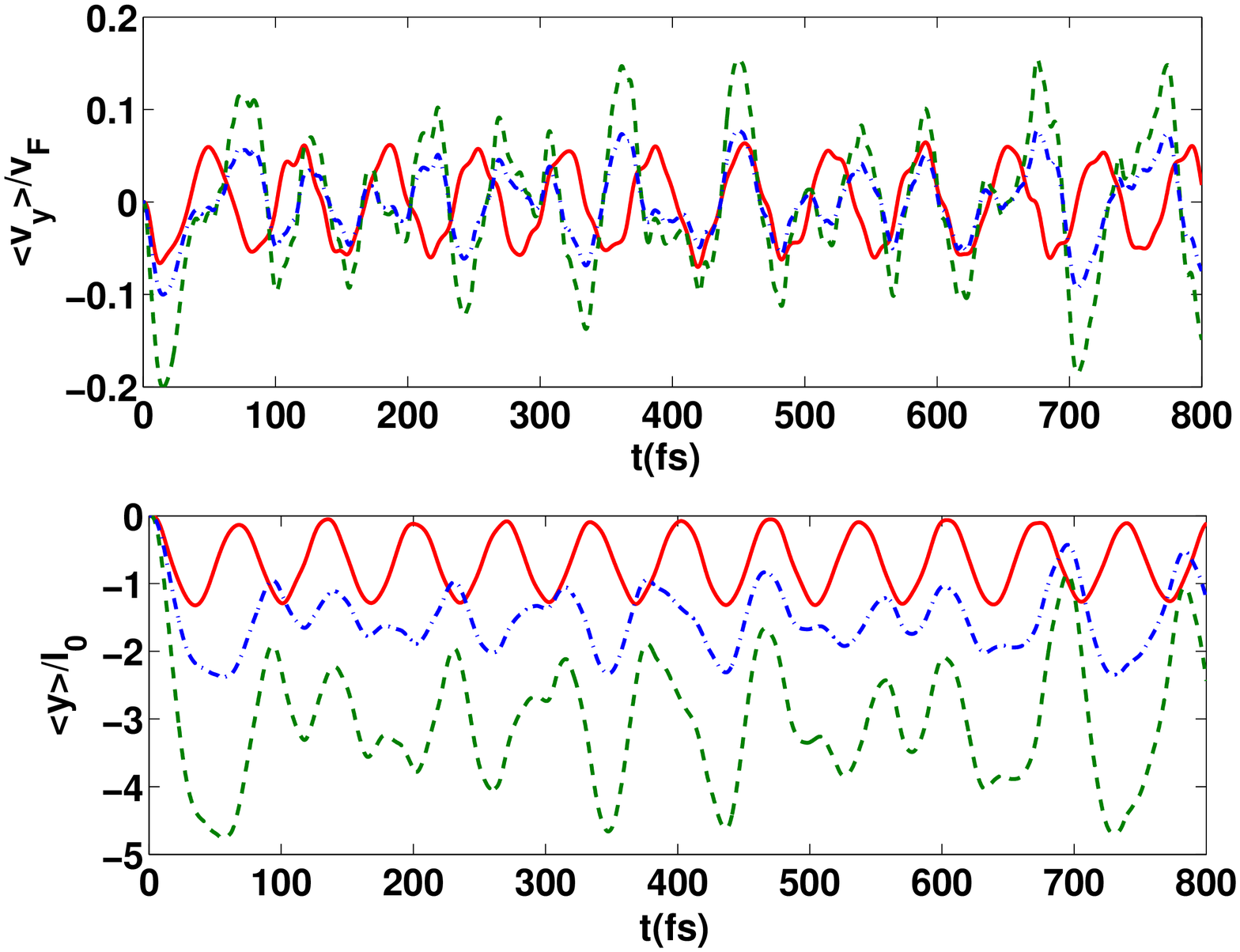}
\caption{Time dependence of the expectation values of $y$-component of velocity and position operators when the 
initial pseudospin was polarized along $x$-direction. Here, solid(red), dashed(green), and 
dot-dashed(blue) lines correspond to different values of $(c_1,c_2,c_3)$, namely, 
$(\sin\varphi, 0, -\cos\varphi)$, $(\cos\varphi, 1, \sin\varphi)$, and $(-\cos\varphi, 1, -\sin\varphi)$,
respectively. We also consider $k_{0x}$=$2\times 10^8$ m$^{-1}$, $l_0$=$8.11$ nm, and $\alpha$=$0.5$.}
\end{center}
\end{figure}

\begin{center}
 \textbf{3.~~{\textit{y}}-polarization}
\end{center}
Finally, we depict the time dependence of the expectation values of position and velocity operators in 
Fig. 12 by considering the initial pseudospin polarization was along $y$-direction. We have the following
choices of $(c_1,c_2,c_3)$ such as $(\sin\varphi, 0, \cos\varphi)$, $(-i\cos\varphi, 1, i\sin\varphi)$,
and $(i\cos\varphi, 1, -i\sin\varphi)$. We take same parameter values as considered for choice 2. Similar to choice 1,
we obtain $\la x(t)\ra=0$ and $\la v_x(t)\ra=0$. For 
all possibilities of $(c_1,c_2,c_3)$, complicated irregular oscillatory patterns are obtained in both $\la y(t)\ra$
and $\la v_y(t)\ra$.

\begin{figure}[h!]
\begin{center}\leavevmode
\includegraphics[width=125mm, height=80mm]{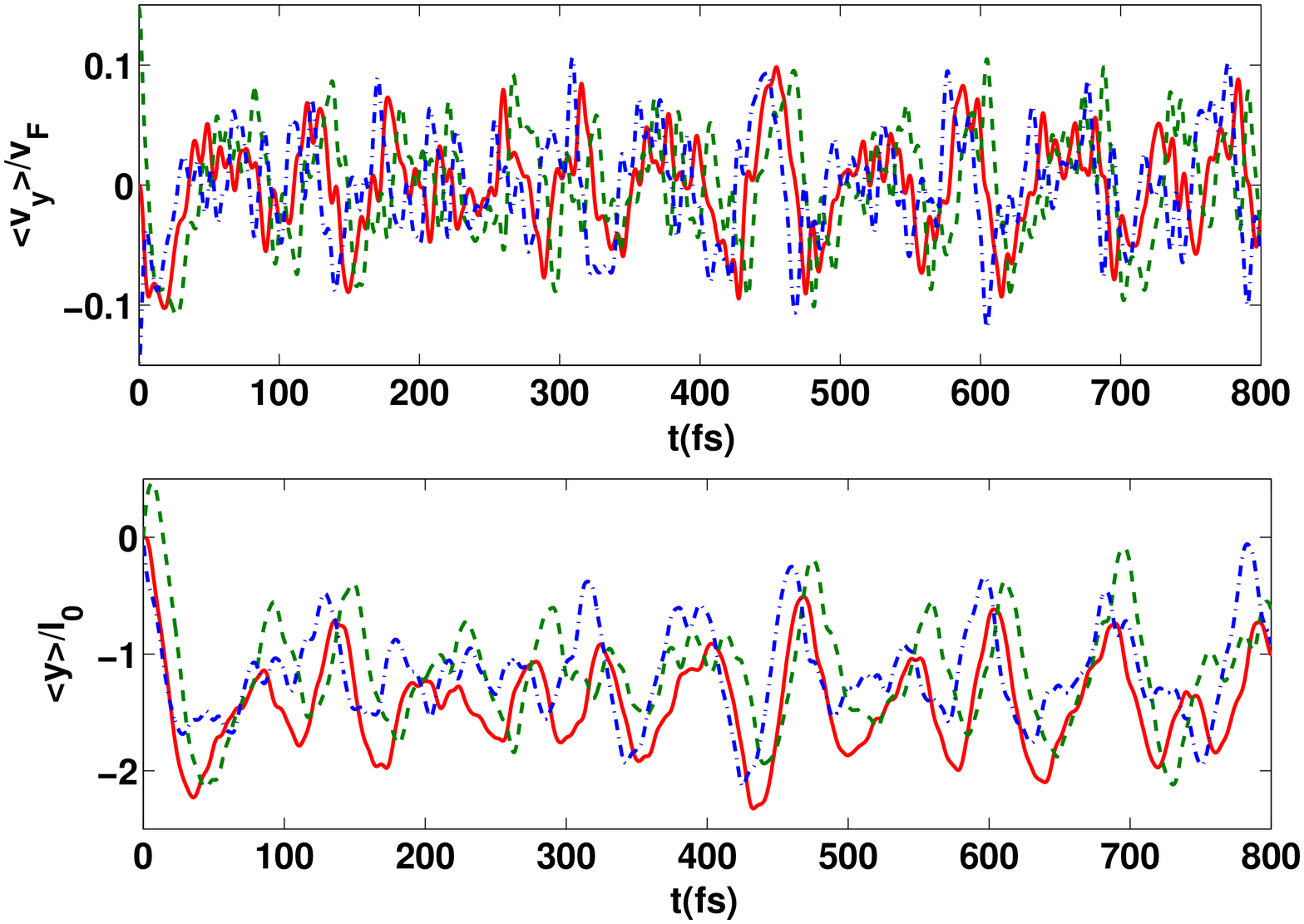}
\caption{Time dependence of the expectation values of $y$-component of velocity and position operators when the 
initial pseudospin was polarized along $y$-direction. Here, solid(red), dashed(green), and 
dot-dashed(blue) lines correspond to different values of $(c_1,c_2,c_3)$, namely, 
$(\sin\varphi, 0, \cos\varphi)$, $(-i\cos\varphi, 1, i\sin\varphi)$, and $(i\cos\varphi, 1, -i\sin\varphi)$,
respectively. We also consider $k_{0x}$=$2\times 10^8$ m$^{-1}$, $l_0$=$8.11$ nm, and $\alpha$=$0.5$.}
\end{center}
\end{figure}

\begin{figure}[t]
\begin{center}\leavevmode
\includegraphics[width=110mm, height=50mm]{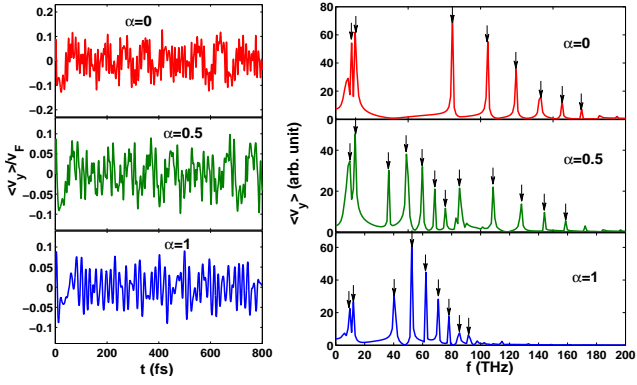}
\caption{The Left panel shows the time dependence of $\la v_y(t) \ra$ for $z$-polarized wave packet
with components $c_1$=$1$, $c_2$=$0$, and $c_3$=$0$. The right panel displays corresponding Fourier
transforms of $\la v_y(t)\ra$. The downward arrows depict the frequency values.
We consider $k_{0x}$=$2\times 10^8$ m$^{-1}$ and  $l_0$=$8.11$ nm.}
\end{center}
\end{figure}

\begin{table}[ht]
 \centering
\begin{tabular}{ |p{1.5cm}| p{1.5cm}| p{1.5cm}| p{1.5cm}|}
\hline
Freq. (THz) & $\alpha$=$0$ & $\alpha$=0.5 & $\alpha$=1\\
\hline
$f_1$& $10.99$ & $9.76$ & $9.76$ \\
$f_2$& $13.43$  & $13.43$ & $12.21$\\
$f_3$& $80.59$ & $36.63$ & $40.29$\\
$f_4$& $105.00$& $48.84$ & $52.50$\\
$f_5$& $124.50$ & $59.83$  & $62.27$\\
$f_6$& $141.60$ & $68.38$ & $70.82$\\
$f_7$& $156.30$  & $75.70$ & $78.14$\\
$f_8$ & $169.70$  & $85.47$ & $85.47$\\
$f_9$ & $-$ & $108.70$ & $91.58$\\
$f_{10}$ & $-$& $128.20$ & $-$\\
$f_{11}$ & $-$ & $144.10$ & $-$\\
$f_{12}$ & $-$& $158.70$ & $-$\\  
\hline
\end{tabular}
\caption{Frequency involved in ZB for different values of $\alpha$ obtained from Fig. 13.}
\end{table}

\subsection{Determination of frequencies involved in ZB}
It would be interesting to find out the frequency components
which are present in the complicated structure of ZB.
As an example we consider the case in which the wave packet was polarized initially along 
$z$-direction with components $c_1$=$1$, $c_2$=$0$, and $c_3$=$0$. The 
time dependence of $\la v_y(t)\ra$ and $\la y(t) \ra$ are already 
shown in Fig. 7.
We make fast Fourier transformation of $\la v_y\ra$ vs $t$ data
to find the frequencies involved in ZB. The corresponding results are shown in Fig. 13.
The arrows in the right panels of Fig. 13 denote the 
frequencies involved in ZB. The revealed frequencies corresponding to  different values of $\alpha$
are given in Table I. Note that the number of frequency components depends on $\alpha$ significantly.
We find, approximately, $8$, $12$, and 
$9$ frequencies for $\alpha$=$0$, $\alpha$=$0.5$, and $\alpha$=$1$, respectively. These frequencies
are governed by all possible 
differences between Landau energy levels. 
In a similar way one can also find out the frequencies corresponding to the 
other choices of $c_1$, $c_2$, and $c_3$.

\section{Summary}
In summary, we have studied ZB of a Gaussian wave packet which represents a quasiparticle in $\alpha$-T$_3$ model.
We also consider the effect of an external transverse magnetic field on ZB. The manifestation of ZB of wave packet
is shown in the expectation values of physical observables like position and velocity.
For zero magnetic field case, 
we find that the ZBs appeared in position and velocity diminish with time.
The problem studied in this article is an example of two frequency ZB for a finite values of 
$\alpha$ e.g. $0<\alpha<1$.
One frequency is originating due to the interference between 
conduction and valence band whereas the other frequency is a
result of interference between either conduction and flat band 
or flat and valence band. It is revealed that ZB depends significantly on 
the nature of the initial pseudospin polarization. Specifically, the 
case with initial pseudospin polarization along $z$-direction is more 
interesting. By considering this particular spin polarization we find that ZB consists of two aforesaid 
frequencies for $0<\alpha<1$ when the initial wave packet was completely located
in any of the $rim$ sites. A transition from $2\Omega_q$-frequency ZB to $\Omega_q$-frequency ZB is unveiled
as $\alpha$ is varied from $0$ to $1$. On contrary, the existence of a single $2\Omega_q$-frequency ZB is realized
for a finite $\alpha$ in the case of initial wave packet being situated in the $hub$ site. The 
timescales over which the ZB persists can be extracted from the approximate results of expectation values
obtained in the large width limit of the wave packet. Other choices of initial pseudospin polarization have produced
some interesting features. In presence of a finite magnetic field the ZB displays complicated permanent oscillations 
as a result of interference among large number of Landau levels. Similar to zero magnetic field case the 
oscillatory pattern depends on the type of initial pseudospin polarization.

\end{document}